\documentclass[showpacs,preprintnumbers,amsmath,amssymb]{revtex4}
\usepackage{amsmath,amssymb,graphics,epsfig,subfigure}
\usepackage{color}

\begin{document}
\renewcommand{\baselinestretch}{1.3}

\title{Null geodesics, quasinormal modes, and thermodynamic phase transition for charged black holes in asymptotically flat and dS spacetimes}

\author{Shao-Wen Wei$^{1,2}$ \footnote{weishw@lzu.edu.cn},
Yu-Xiao Liu$^{1}$ \footnote{liuyx@lzu.edu.cn},}

\affiliation{ $^{1}$Institute of Theoretical Physics $\&$ Research Center of Gravitation, Lanzhou University, Lanzhou 730000, People's Republic of China,\\
$^{2}$Department of Physics and Astronomy, University of Waterloo, Waterloo, Ontario, Canada, N2L 3G1}

\begin{abstract}
The numerical study indicates that there exists a relation between the quasinormal modes and the Davies point for a black hole. In this paper, we analytically study this relation for the charged Reissner-Nordstr\"{o}m black holes in asymptotically flat and dS spacetimes. In the eikonal limit, the angular velocity $\Omega$ and the Lyapunov exponent $\lambda$ of the photon sphere, respectively, corresponding to the real and imaginary parts of the quasinormal modes are obtained from the null geodesics. Both in asymptotically flat and dS spacetimes, we observe the spiral-like shapes in the complex quasinormal mode plane. However, the starting point of the shapes do not coincide with the Davies point. Nevertheless, we find a new relation that the Davies point exactly meet the maximum of the temperature $T$ in the $T$-$\Omega$ and $T$-$\lambda$ planes. In higher dimensional asymptotically flat spacetime, even there is no the spiral-like shape, such relation still holds. Therefore, we provide a new relation between the black hole thermodynamics and dynamics. Applying this relation, we can test the black hole thermodynamic property by the quasinormal modes.
\end{abstract}



\maketitle
\section{Introduction}

Recently, the direct observation of the gravitational waves by the LIGO-Virgo collaboration marks the coming of the new era of astronomical observation~\cite{Abbott}. With the data of the gravitational waves, the parameters of the black hole, such as the mass and spin, are determined. It is expected that with the improving of the detection precision, more details about the black holes and the gravity theories will be revealed.

It is widely known that, in the last stage of the black hole merger, the ring down modes can be well described by the quasinormal modes (QNMs). These modes are complex numbers, and can be regarded as the perturbations of the black hole in a given spacetime. In order to probe the characteristic properties of the black hole by using the gravitational waves, one first needs to know what is directly linked to the gravitational waves or the QNMs.

On the other hand, after the discovery of the four laws~\cite{Hawking,Bekensteina,Bekensteinb,Bardeen}, black hole thermodynamics continues to be one of the most interesting subjects in the black hole physics. For a Schwarzschild black hole, it has only one thermodynamic phase of negative heat capacity, which indicates that the isolated black hole is thermodynamic unstable. However, when the black hole gets other hairs, such as the electric charge and spin, besides the negative heat phase, there will present a phase of positive heat capacity. Across these two black hole phases, the sign of the heat capacity of the black hole changes. More intriguing thing is that the heat capacity goes to infinity at the joined point of these two phases. This behavior of the heat capacity may indicate a phase transition of the black hole system between the thermodynamic unstable and stable phases. It was found by Davies \cite{Daviespc,Daviespcb,Daviespcc} that this second order thermodynamic phase transition just takes place at the singular point of the heat capacity.

Therefore, the study of the relation between the phase transition and QNMs gets a first step on understanding black hole thermodynamics from black hole dynamics. The issue was studied in Refs. \cite{Koutsoumbas,Koutsoumbasb,Shen,Rao}. Subsequently, Jing and Pan~\cite{Jing} proposed a clear relation between them. It states that, for a given quantum number beyond some certain critical value, the QNMs of the Reissner-Nordstr\"{o}m (RN) black hole start to get a spiral-like shape in the complex QNM plane when the Davies point is approached. They found that both the real and imaginary parts of the QNMs behave as oscillatory functions of the black hole charge. However, Berti and Cardoso~\cite{Berti} argued that this result is probably a numerical coincidence due to the fact that it cannot be generalized to other black hole backgrounds. While, He et al.~\cite{He} applied the study to the charged Kaluza-Klein black hole with squashed horizon. It was found that there exists the spiral-like shape, and its starting point is consistent with the Davies point within 8\%, or even lower for some other cases. These results imply that the relation between the black hole thermodynamics and dynamics perhaps is non-trivial. Some other works can be found in Refs. \cite{He2,LinLin,Sup}

Due to the complexity of the QNMs, only the numerical result is available, and an analytical check for the relation is still lacked. Actually, according to the light ring/QNMs correspondence~\cite{Cardoso}, the QNMs are allowed to be parametrized by the radius of the photon sphere in spherically symmetric spacetime, or the radius of the light ring in stationary spacetime. Moreover, this method is only effective in the eikonal limit, where the quantum numbers must have large values. What interesting is that this condition naturally satisfies the requirement of Ref.~\cite{Jing} that the quantum numbers must be larger than some critical values. Thus by making use this light ring/QNMs correspondence, we can analytically check the relation given in Ref.~\cite{Jing}, and this is the main purpose of this paper. Several years ago, this correspondence was also applied to establish a universal relation between the QNMs and the black hole lensing for asymptotically flat black holes with or without spin~\cite{Stefanov,Weic}.

In fact, we have carried out the study on exploring the relation between the photon sphere (light ring) and the Van der Waals type phase transition~\cite{Kubiznak} for the charged or rotating black hole in AdS spacetime~\cite{Weia,Weib}. For the charged AdS black hole, it was found that there are the non-monotonic behaviors of the photon sphere radius and the minimum impact parameter when the pressure and temperature are below their critical values, while the behaviors disappear when the parameters excess their critical values~\cite{Weia}. So this suggests that such non-monotonic behaviors can reflect the small/large black hole phase transition. Among the small/large black hole phase transition, the radius and the minimum impact parameter of the photon sphere have sudden changes, and these changes can serve as order parameters to describe the small/large black hole phase transition. More interestingly, there exists a universal critical exponent of $1/2$ for the changes of the radius and the minimum impact parameter near the critical point, and this exponent is also independent of the dimension of the spacetime~\cite{Weia}. Such study was also generalized to the rotating Kerr-AdS black hole~\cite{Weib} and the Born-Infeld-AdS black hole~\cite{Weid}. Besides, one more new issue was examined, and the result shows that the temperature and pressure corresponding to the extremal points of the radius or the angular momentum of the light rings exactly agree with that of the metastable curves from the thermodynamic side. All the study confirms that there exists a relation between the null geodesics and small/large black hole phase transition in the AdS spacetime. The correspondence of the phase transition and the time-like geodesics can also be found in Ref.~\cite{Bhamidipati}.

Since the angular velocity and Lyapunov exponent of the photon sphere are, respectively, corresponded to the real and imaginary parts of the QNMs, we mainly, in this paper, aim to explore the relation between the angular velocity, Lyapunov exponent and the Davies phase transition point in a static and spherically symmetric spacetime. We expect our study on this relation could provide a possible way to test the black hole thermodynamics with the observation of the gravitational waves.

The paper is organized as follows. In Sec. \ref{null}, we briefly review the null geodesics and photon sphere for the black hole in a static and spherically symmetric spacetime. In Sec. \ref{RNn}, we calculate the Davies point and the photon sphere from the null geodesics for a four-dimensional charged RN black hole. Then we analytically obtain the angular velocity and Lyapunov exponent, and further explore the relation between the QNMs and the Davies point proposed in Ref.~\cite{Jing}. But unfortunately, the relation does not exactly hold. However, we propose a new and exact relation between the Davies point and the maximum of the temperature. Then we extend the study to higher dimensional spacetime in Sec. \ref{higherdime}. It is surprise that the spiral-like shape does not appear in higher spacetime. Nevertheless, our new relation is still effective. In Sec. \ref{RNdsBH}, we also apply the study to the charged RN black hole in dS spacetime and find some novel results.
Finally, the conclusions are presented in Sec. \ref{Conclusion}.

\section{Null geodesics and photon sphere}
\label{null}

In $d(\geq 4)$-dimensional, static, and spherically symmetric spacetime, a black hole can be described by the following metric
\begin{eqnarray}
 ds^{2}&=&-f(r)dt^{2}+\frac{1}{f(r)}dr^{2}+r^{2}d\Omega_{(d-2)}^{2},\label{metric}
\end{eqnarray}
where $d\Omega_{(d-2)}^{2}$ is the line element on a unit $(d-2)$-dimensional sphere $S^{(d-2)}$. The usual angular coordinates $\theta_{i}\in [0,\;\pi]$ $(i=1,\cdots,d-3)$ and $\phi\in [0,\;2\pi]$. The metric function $f(r)$ depends on the radial coordinate $r$ and other black hole parameters, such as the mass and charge.

Next, we would like to study the motion of a free photon in the black hole background. Since the spactime we focus on is spherically symmetric, without loss of generality, one can consider the motion limited in the equatorial hyperplane ($\theta_{i}=\frac{\pi}{2}$ for $i=1,\cdots ,d-3$). Then the Lagrangian for a photon reads
\begin{eqnarray}
 2\mathcal{L}=-f(r)\dot{t}^{2}+\dot{r}^{2}/f(r)+r^{2}\dot{\phi}^{2}.
 \label{lagrangian}
\end{eqnarray}
The dot over a symbol denotes the ordinary differentiation with respect to an affine parameter. With the help of this Lagrangian, the generalized momentum defined by $p_{\mu}=\frac{\partial \mathcal{L}}{\partial \dot{x}^{\mu}}=g_{\mu\nu}\dot{x}^{\nu}$ has the following form
\begin{eqnarray}
 p_{t}   &=&-f(r)\dot{t}\equiv -E,\label{pt}\\
 p_{\phi}&=&r^{2}\dot{\phi}\equiv l,\label{phi}\\
 p_{r}   &=&\dot{r}/f(r).
\end{eqnarray}
For this spacetime, there are two Killing fields $\partial_{t}$ and $\partial_{\phi}$. Thus there are two conservation constants $E$ and $l$ corresponded to each geodesics, which, in fact, are the energy and orbital angular momentum of the photon, respectively. The $t$-motion and $\phi$-motion can be easily obtained by solving Eqs.~(\ref{pt}) and (\ref{phi}),
\begin{eqnarray}
 \dot{t}&=&\frac{E}{f(r)},\\
 \dot{\phi}&=&\frac{l}{r^{2}}.\label{phit2}
\end{eqnarray}
The Hamiltonian for this system reads
\begin{eqnarray}
 2\mathcal{H}&=&2(p_{\mu}\dot{x}^{\mu}-\mathcal{L})\nonumber\\
     &=&-f(r)\dot{t}^{2}+\dot{r}^{2}/f(r)+r^{2}\dot{\phi}^{2}\nonumber\\
     &=&-E\dot{t}+l\dot{\phi}+\dot{r}^{2}/f(r)=0.
\end{eqnarray}
Using the $t$-motion and $\phi$-motion, it is easy to get the radial $r$-motion, which can be expressed in the following form
\begin{eqnarray}
 \dot{r}^{2}+V_{\rm eff}=0,\label{rt2}
\end{eqnarray}
where the effective potential is
\begin{eqnarray}
 V_{\rm eff}=\frac{l^{2}}{r^{2}}f(r)-E^{2}.\label{veff}
\end{eqnarray}
Since $\dot{r}^{2}>$0, the photon is only allowed to appear at the region of negative potential. For a photon coming from infinity, it will fall into the black hole if it has low angular momentum. While for the larger angular momentum case, it will be bounded back to infinity when it meets a turning point determined by $V_{\rm eff}=0$. Among these two cases, there exists a critical case that the photon traveling from infinity plunges into a circular orbit and will orbit the black hole one loop by one loop. Nevertheless, such orbit is unstable. Due to the spherically symmetry of the spacetime, such circular orbit is a sphere and known as the photon sphere. The conditions to determine this photon sphere are
\begin{eqnarray}
 V_{\rm eff}=0,\quad \frac{\partial V_{\rm eff}}{\partial r}=0,\quad \frac{\partial^{2} V_{\rm eff}}{\partial r^{2}}<0.
\end{eqnarray}
The first two conditions can determine the radius $r_{\rm ps}$ and the critical angular momentum $l_{\rm ps}$ of the photon sphere. The third one guarantees that the photon sphere is unstable.

Substituting the effective potential (\ref{veff}) into the second condition, we obtain the equation that the radius $r_{\rm ps}$ must satisfy
\begin{eqnarray}
 2f(r_{\rm ps})-r_{\rm ps}\partial_{r}f(r_{\rm ps})=0.\label{rps1}
\end{eqnarray}
For a given metric function $f(r)$, we can obtain $r_{\rm ps}$ by solving it. Then from the first condition, the critical angular momentum can be obtained as
\begin{eqnarray}
 u_{\rm ps}=\frac{l_{\rm ps}}{E}=\frac{r}{\sqrt{f(r)}}\bigg|_{r_{\rm ps}}.\label{uspp}
\end{eqnarray}
The third condition is also closely linked to the QNMs of the black hole. Without loss of generality, we set $E$=1.

Now, let us turn to the QNMs. In the eikonal limit ($l\gg1$), the QNMs $\omega_{\rm Q}$ can be calculated with the property of the photon sphere~\cite{Cardoso,Goebel,Mashhoon}
\begin{eqnarray}
 \omega_{\rm Q}=l\Omega-i\left(n+\frac{1}{2}\right)|\lambda|,
\end{eqnarray}
where $n$ and $l$ are, respectively, the number of the overtone and the angular momentum of the perturbation. The other two quantities $\Omega$ and $\lambda$ are the angular velocity and Lyapunov exponent of the photon sphere, which can be parametrized by the null geodesics
\begin{eqnarray}
 \Omega=\frac{\dot{\phi}}{\dot{t}}\bigg|_{r_{\rm ps}},\quad
 \lambda=\sqrt{-\frac{V_{\rm eff}''}{2\dot{t}^{2}}}\bigg|_{r_{\rm ps}}.
\end{eqnarray}
Note that $V_{\rm eff}''<0$, the term under the square root is positive. In the background of (\ref{metric}), these two quantities can be expressed as
\begin{eqnarray}
 \Omega=\frac{\sqrt{f_{\rm ps}}}{r_{\rm ps}}=\frac{1}{l_{\rm ps}},\quad
 \lambda=\sqrt{\frac{f_{\rm ps}(2f_{\rm ps}-r^{2}_{\rm ps}f_{\rm ps}'')}{2r_{\rm ps}^{2}}},\label{omelam}
\end{eqnarray}
where we have used (\ref{rps1}). For a given metric function, one can easily obtain $\Omega$ and $\lambda$. Taking Schwarzschild black hole as an example, we have $r_{\rm ps}=3M$ and $l_{\rm ps}=3\sqrt{3}M$, which gives $\Omega=1/3\sqrt{3}M$ and $\lambda=1/3\sqrt{3}M$. Although this method is effective for large $l$, it is still accurate for some cases with small $l$~\cite{Iyer,Bertib}.

Following Refs.~\cite{Cardoso,Decanini}, we would like to give a brief review on the effective application of the above treatment. It is well known that, in the eikonal limit, the scalar, electromagnetic and gravitational perturbations in the background (\ref{metric}) have the same behavior, so we take the scalar perturbation as an example. It obeys the Klein-Gordon equation. Taking a separate method, the radial partial wave function $\Phi_{l}(r)$ satisfies the Regge-Wheeler equation
\begin{eqnarray}
 \frac{d^{2}\Phi_{l}}{dr_{*}^{2}}-V_{l}\Phi_{l}=0,
\end{eqnarray}
where the convenient ``tortoise" coordinate $r_{*}$ ranges from $-\infty$ to $+\infty$. The potential $V_{l}(r)$ reads
\begin{eqnarray}
 V_{l}(r)=\frac{l(l+d-3)}{r^{2}}f(r)+\frac{(d-2)(d-4)}{4r^{2}}f^{2}(r)+\frac{d-2}{2r}f(r)f'(r)-\omega^{2},\label{lele}
\end{eqnarray}
where $\omega$ is the frequency of the perturbation and can be corresponded to the energy $E=\hbar \omega$ of the perturbation particle. In the eikonal limit $l\gg1$, the second and third terms in the right hand  side of (\ref{lele}) can be ignored, and $l(l+d-3)\approx l^{2}$, thus the potential reduces to
\begin{eqnarray}
 V_{l}(r)=\frac{l^{2}}{r^{2}}f(r)-\omega^{2}.\label{veffg}
\end{eqnarray}
It is obvious that these two potentials (\ref{veff}) and (\ref{veffg}) are the same, as well as the equation of  motion. Therefore, it is reasonable to obtain the QNMs following the null geodesics method. One thing worth to note is that this method cannot be extended to asymptotically AdS spacetime.

\section{Four-dimensional charged Reissner-Nordstr\"{o}m black holes}
\label{RNn}

\subsection{Thermodynamics and Davies point}

\begin{figure}
\center{\includegraphics[width=6cm]{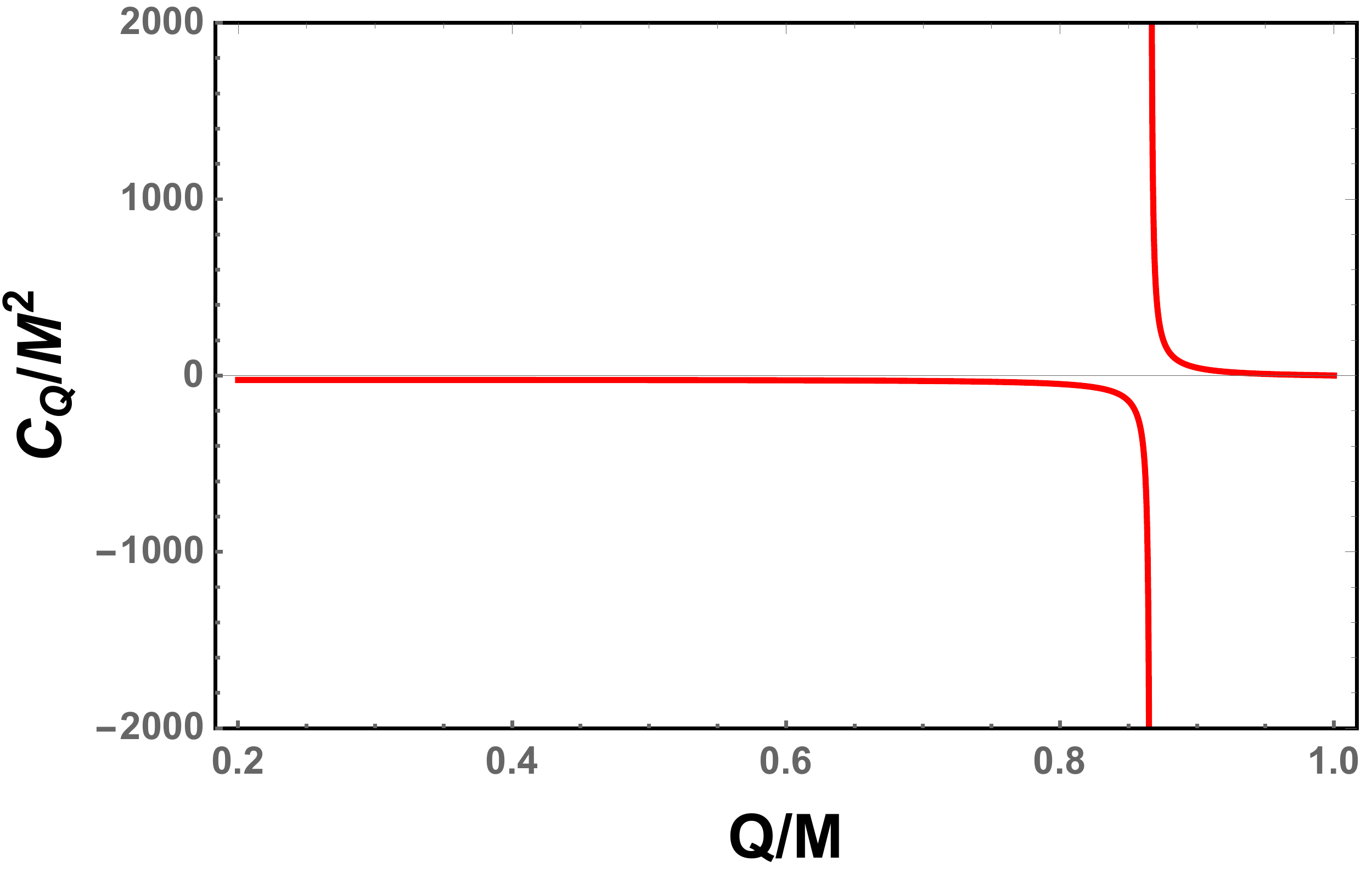}}
\caption{The heat capacity $C_{Q}$ vs. $Q$ for the four-dimensional charged RN black hole. The heat capacity diverges at $Q_{\rm D}/M=\frac{\sqrt{3}}{2}$.}\label{ptHeatcap}
\end{figure}

The four-dimensional charged RN black hole solution is given by the metric (\ref{metric}) with the following function
\begin{eqnarray}
 f(r)=1-\frac{2M}{r}+\frac{Q^{2}}{r^{2}}.
\end{eqnarray}
The parameters $M$ and $Q$ are the mass and charge of the black hole. Solving $f(r)=0$, one can easily get the outer and inner horizons of the black hole, which are located at
\begin{eqnarray}
 r_{\pm}=M\pm\sqrt{M^{2}-Q^{2}}.
\end{eqnarray}
The entropy and temperature corresponding to the event horizon are
\begin{eqnarray}
 S&=&\pi (M+\sqrt{M^{2}-Q^{2}})^{2},\\
 T&=&\frac{M^{2}-Q^{2}+M\sqrt{M^{2}-Q^{2}}}
   {2\pi(M+\sqrt{M^{2}-Q^{2}})^{3}}.\label{ttemp}
\end{eqnarray}
The heat capacity at a fixed charge is
\begin{eqnarray}
 C_{Q}=T\left(\frac{dS}{dT}\right)_{Q}
           =\frac{2S(S-\pi Q^{2})}{3\pi Q^{2}-S}
           =\frac{2\pi(M+\sqrt{M^{2}-Q^{2}})^{2}((M+\sqrt{M^{2}-Q^{2}})^{2}-Q^{2})}{3 Q^{2}-(M+\sqrt{M^{2}-Q^{2}})^{2}}.
\end{eqnarray}
We plot the heat capacity $C_{Q}$ in Fig.~\ref{ptHeatcap}. It is clearly that for mall charge case, $C_{Q}$ is negative just like the Schwarzschild black hole, indicating that the black hole is thermodynamic unstable. While for the large charge case, $C_{Q}$ becomes positive and the black hole is thermodynamic stable. Among these two cases, the heat capacity diverges at
\begin{eqnarray}
 Q_{\rm D}=\frac{\sqrt{3}}{2}M\approx0.8660M,\label{Dav}
\end{eqnarray}
which is called the Davies point. It was shown by Davies~\cite{Daviespc,Daviespcb,Daviespcc} that this divergence point measures a phase transition of the black hole between the thermodynamic unstable and stable phases.

\subsection{Quasinormal modes and Davies point}

In Ref.~\cite{Jing}, Jing and Pan explored the Davies point by making use of the QNMs. They proposed that when a black hole passes through this phase transition point, the behavior of the QNMs in the complex $\omega_{Q}$ plane starts to have a spiral-like shape, and both the real and imaginary parts of the QNMs for a given overtone number and angular quantum number beyond the critical values become the oscillatory functions of the black hole charge. This provides an interesting dynamic probe to the thermodynamic phase transition. However, in Ref.~\cite{Cardoso}, Cardoso et al. argued that this result is probably a numerical coincidence. Nevertheless, this relation between the thermodynamic and dynamical properties of black holes was further confirmed in Ref.~\cite{He}. In the following, we would like to analytically examine this relation in detail.

\begin{figure}
\center{\includegraphics[width=6cm]{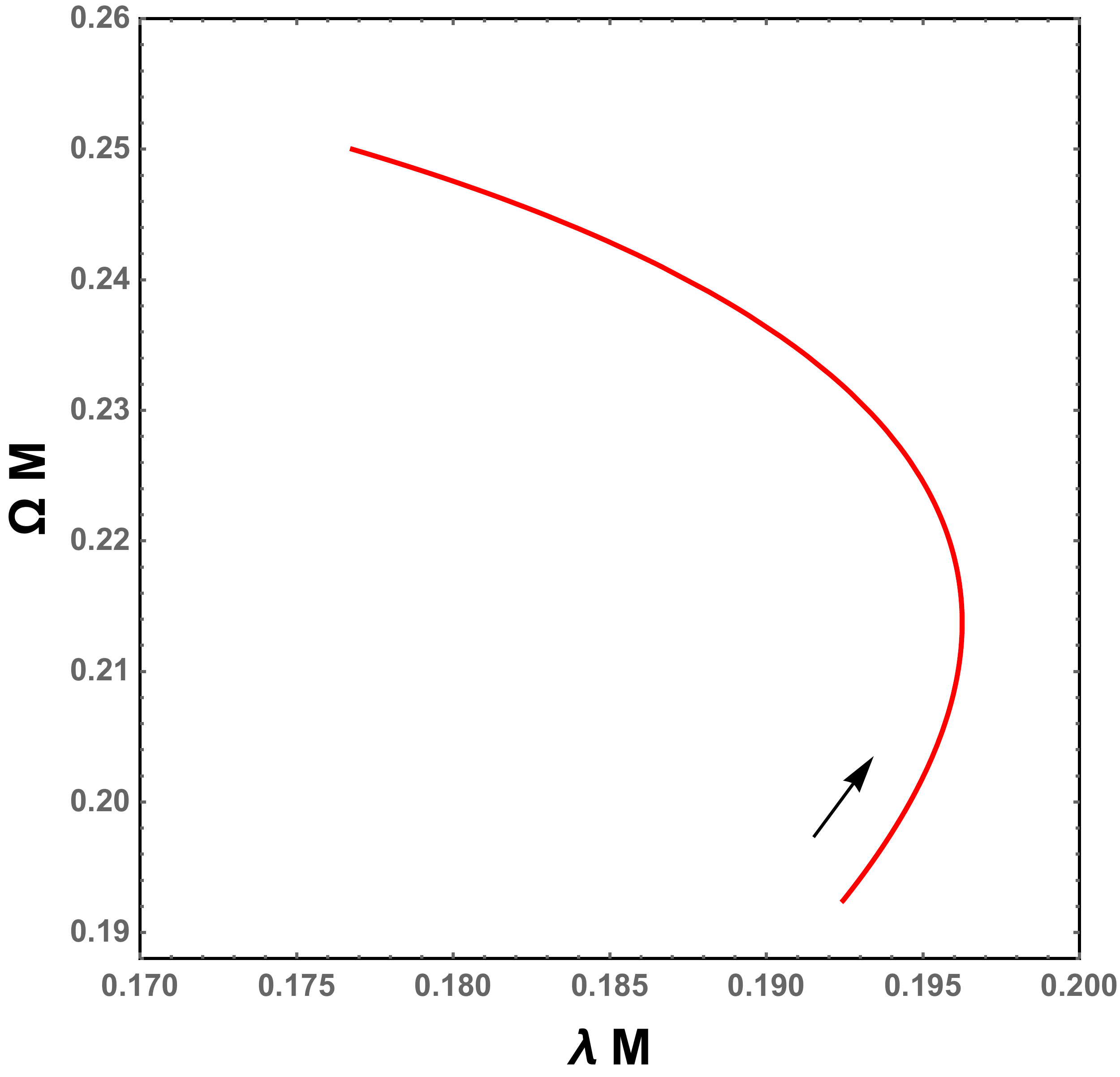}}
\caption{The angular velocity and Lyapunov exponent in the $\Omega$-$\lambda$ plane. The black arrow indicates the increase of the charge $Q$.}\label{ptmegaLam}
\end{figure}

\begin{figure}
\center{\includegraphics[width=6cm]{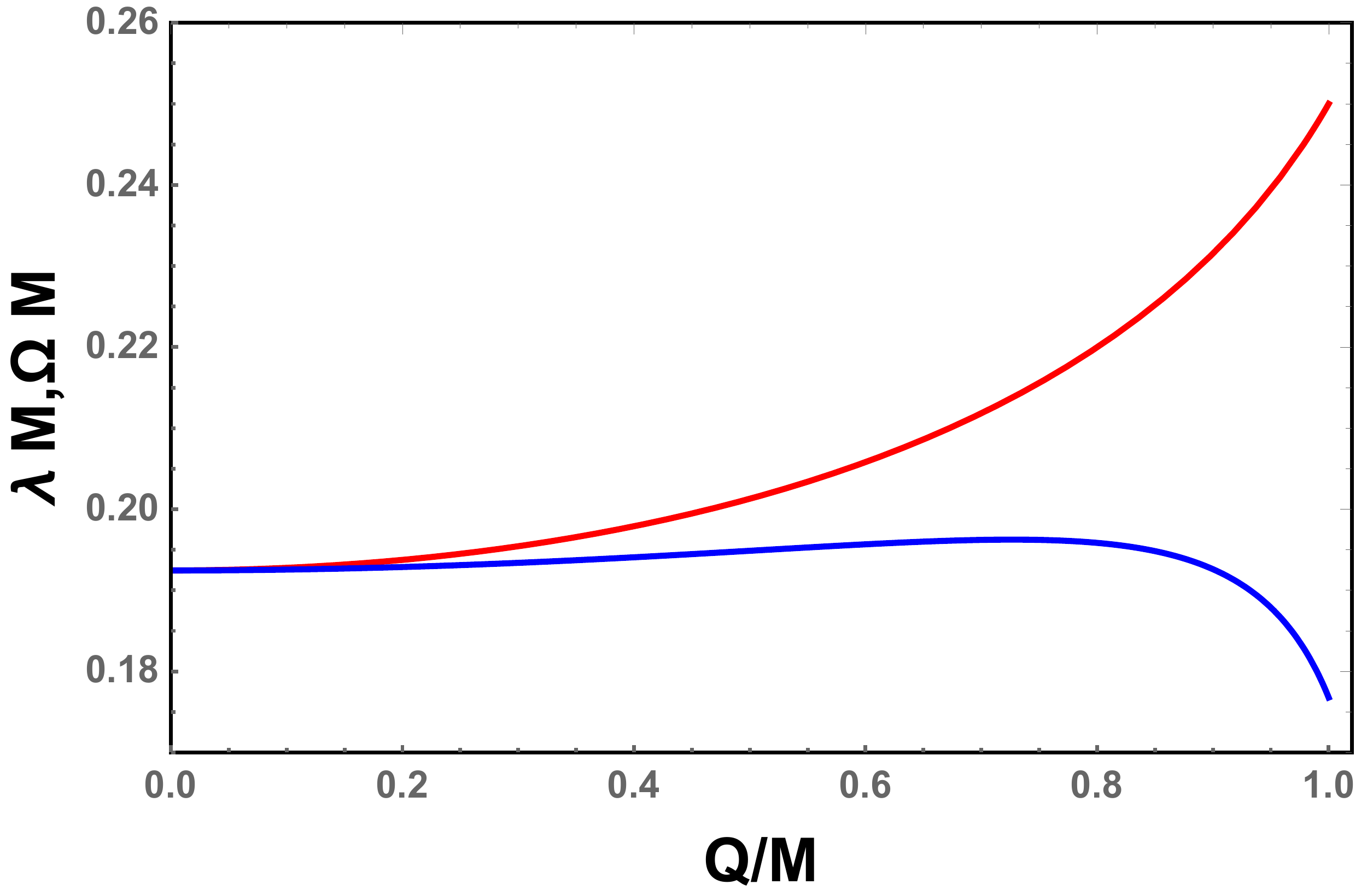}}
\caption{The angular velocity $\Omega$ (top red line) and the Lyapunov exponent $\lambda$ (bottom blue line) as functions of the charge $Q$.}\label{ptloq}
\end{figure}

As shown in Ref.~\cite{Jing}, the relation only holds for large overtone number. In our approach, we work in the eikonal limit, $n\sim l\gg1$, and thus this condition is naturally satisfied. For the charged black hole, the radius of the photon sphere can be obtained by solving (\ref{rps1})
\begin{eqnarray}
 r_{\rm ps}=\frac{1}{2} \left(3M+\sqrt{9 M^2-8 Q^2}\right).\label{rpsq}
\end{eqnarray}
For the case of $Q=0$, we have $r_{\rm ps}=3M$ for the Schwarzschild black hole. And for the extremal charged black hole $Q=M$, the photon sphere will locate at $r=2M$. Plugging $r_{\rm ps}$ into (\ref{omelam}), we obtain the angular velocity and Lyapunov exponent of the photon sphere
\begin{eqnarray}
 \Omega&=&\frac{1}{3M+\sqrt{9M^{2}-8Q^{2}}}\sqrt{2+\frac{M(\sqrt{9M^{2}-8Q^{2}}-3M)}{2Q^{2}}},\\
 \lambda&=&\frac{4\sqrt{\left(M(3M+\sqrt{9M^{2}-8Q^{2}})-2Q^{2}\right)\left(3M(3M+\sqrt{9M^{2}-8Q^{2}})-8Q^{2}\right)}}{M\left(3M+\sqrt{9M^{2}-8Q^{2}}\right)^{3}}.
\end{eqnarray}
In the small charge limit, we have
\begin{eqnarray}
 \Omega &\sim& \frac{1}{3\sqrt{3}M}+\frac{Q^{2}}{18\sqrt{3}M^{3}}+\mathcal{O}(Q^{4}),\\
 \lambda &\sim& \frac{1}{3\sqrt{3}M}+\frac{Q^{2}}{54\sqrt{3}M^{3}}+\mathcal{O}(Q^{4}).
\end{eqnarray}
It was shown in Ref.~\cite{Cardoso} that the spiral-like shape exists in the complex $\omega_{Q}$ plane when the Davies point is passed by. Here we would like to examine this result. It is convenient to show it in the $\Omega$-$\lambda$ plane. The result is given in Fig.~\ref{ptmegaLam}. Obviously, with the increase of the charge $Q$, there will appear the spiral-like shape at a certain value of $Q$. For clarity, we plot the angular velocity $\Omega$ (top red line) and Lyapunov exponent $\lambda$ (bottom blue line) as a function of the charge $Q$ in Fig.~\ref{ptloq}. Interestingly, $\Omega$ is a monotonically increasing function of $Q$. However, $\lambda$ first slowly increases with the charge, and approaches its maximum at a certain value of $Q$, then it decreases. So no oscillatory exists for $\Omega$ or $\lambda$, which is different from that of Ref.~\cite{Jing}. Moreover, the spiral-like shape in the $\Omega$-$\lambda$ plane is resulted by the nonmonotonic behavior of $\lambda$. The starting point of the spiral-like shape corresponds to the maximum of $\lambda$. After a simple calculation, we find the point is
\begin{eqnarray}
 Q_{\rm S}=\frac{\sqrt{51-3\sqrt{33}}}{8}M\approx0.7264M.\label{qss}
\end{eqnarray}
So the starting point of the spiral-like shape is smaller than the Davies point given in (\ref{Dav}). The deviation between them is about 16\%.

In summary, for the charged RN black hole, the Davies point and the starting point of the spiral-like shape are not the same. So the relation proposed in Ref.~\cite{Jing} is not a precise one between the thermodynamics and dynamics of the black hole. Nevertheless, the spiral-like shape can reflect the existence of the thermodynamic phase transition or Davies point. Moreover, we note that the black hole with charge $Q_{\rm S}$ (\ref{qss}) is characterized by the fastest relaxation rate among all the charged RN black holes \cite{Hods}.


\subsection{New relation}
\label{newrelation}

As shown above, the relation given in Ref.~\cite{Jing} is an approximate one. Here, we would like to examine whether a new precise relation exists.

Motivated by our recent work~\cite{Weia,Weib}, where we showed that there exists a relation between the black hole's null geodesics and the thermodynamic phase transition of liquid/gas type in AdS spacetime. We expect to apply it to the charged RN black hole to probe the relation between null geodesics and the Davies point.

\begin{figure}
\center{\subfigure[]{\label{TOmega}
\includegraphics[width=6cm]{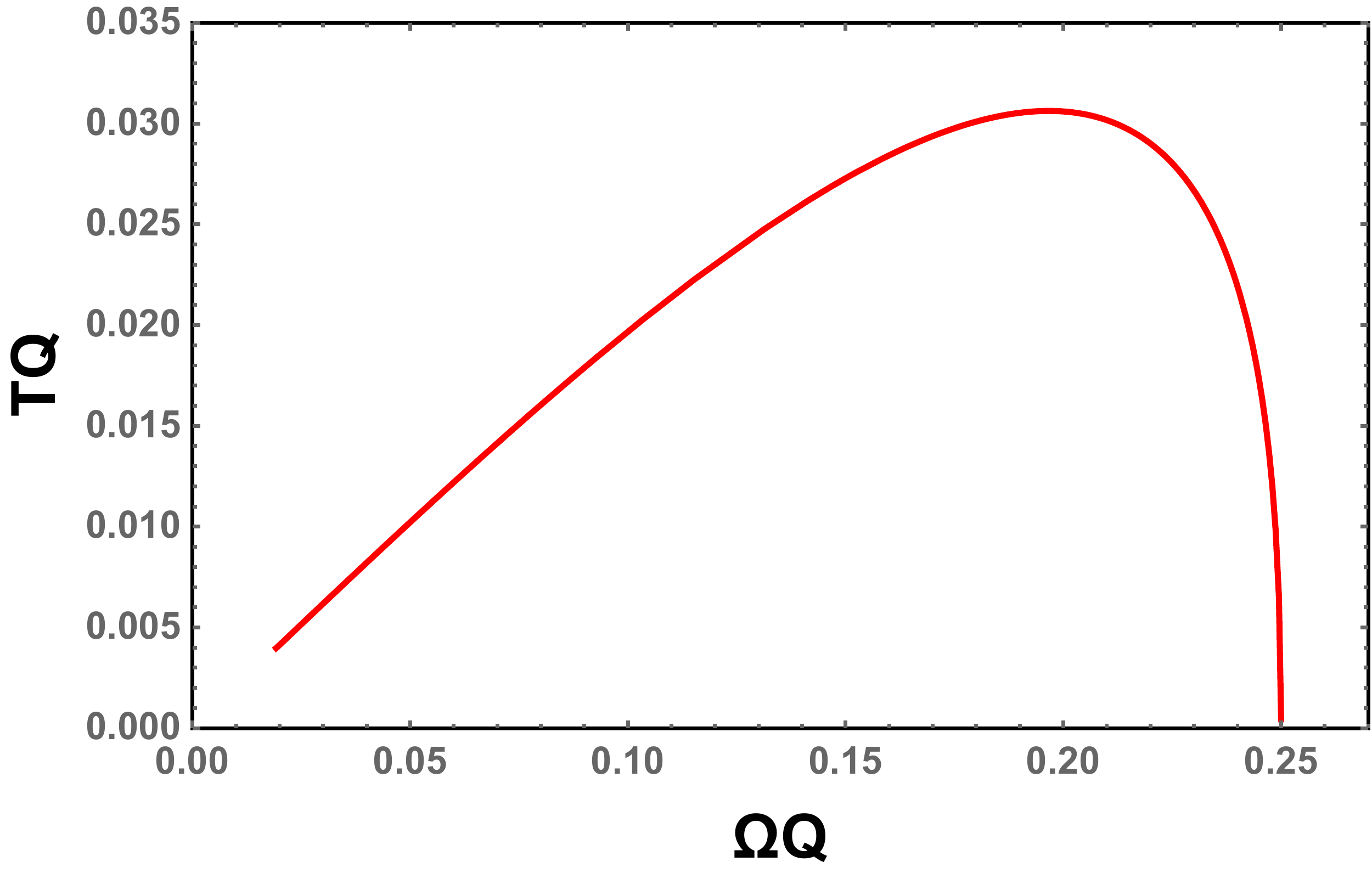}}
\subfigure[]{\label{Tlambda}
\includegraphics[width=6cm]{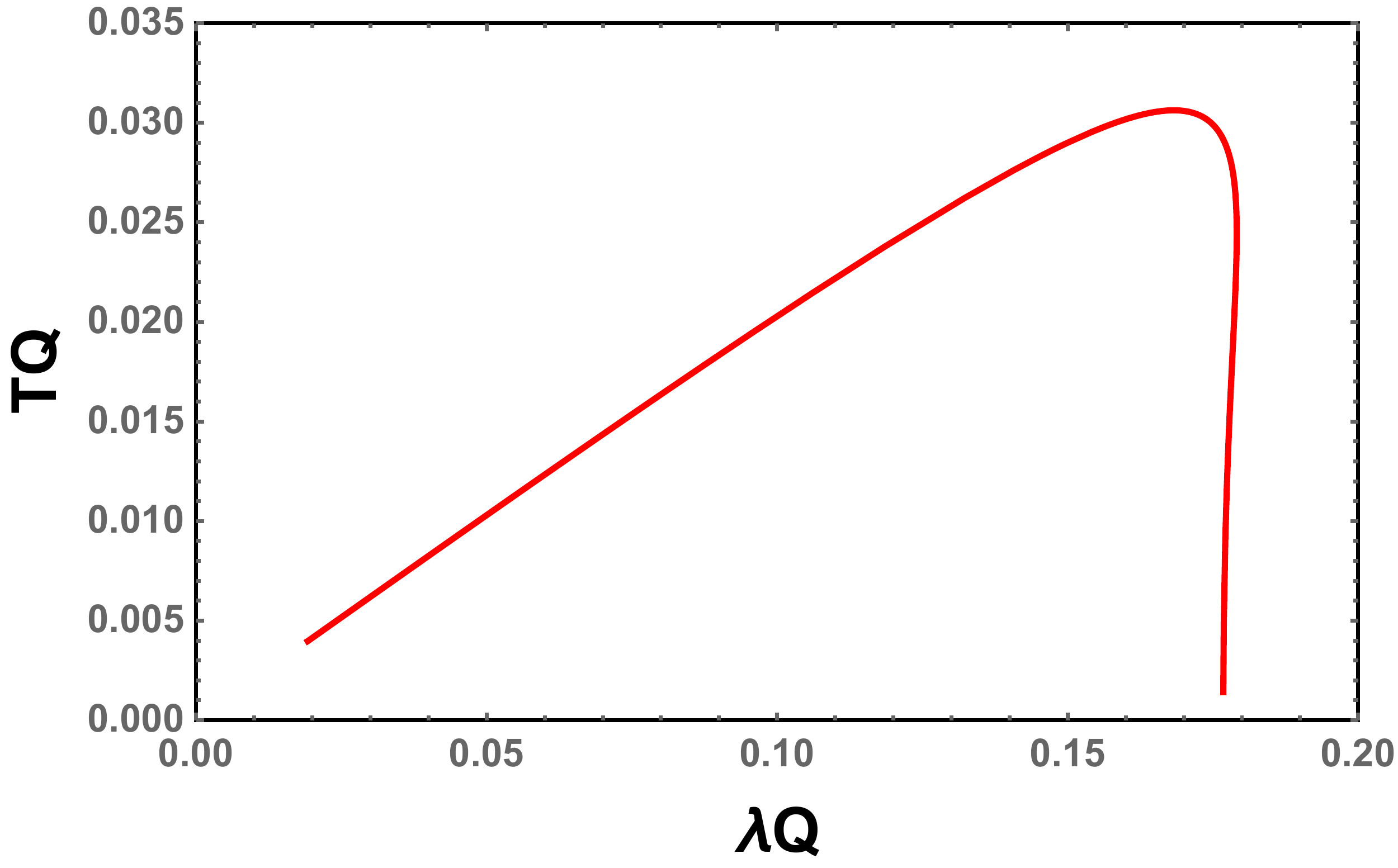}}}
\caption{(a) $T$ vs. $\Omega$. (b) $T$ vs. $\lambda$ for the four-dimensional charged RN black hole.}\label{ppTlambda}
\end{figure}

From (\ref{rpsq}), we can express the black hole mass with the radius of the photon sphere
\begin{eqnarray}
 M=\frac{2Q^{2}+r_{\rm ps}^{2}}{3r_{\rm ps}}.
\end{eqnarray}
Substituting it into (\ref{ttemp}), the temperature will be of the form
\begin{eqnarray}
 T=\frac{3r_{\rm ps}\sqrt{(r_{\rm ps}^{2}-Q^{2})(r_{\rm ps}^{2}-4Q^{2})}\left(\sqrt{(r_{\rm ps}^{2}-Q^{2})(r_{\rm ps}^{2}-4Q^{2})}+(2Q^{2}+r_{\rm ps}^{2})\right)}{2\pi \left(8Q^{4}-Q^{2}r_{\rm ps}^{2}+2r_{\rm ps}^{4}+2(2Q^{2}+r_{\rm ps}^{2})\sqrt{(r_{\rm ps}^{2}-Q^{2})(r_{\rm ps}^{2}-4Q^{2})}\right)^{3/2}}.
\end{eqnarray}
In terms of $r_{\rm ps}$ and $Q$, the angular velocity and the Lyapunov exponent are
\begin{eqnarray}
 \Omega&=&\frac{\sqrt{r_{\rm ps}^{2}-Q^{2}}}{\sqrt{3}r_{\rm ps}^{2}},\\
 \lambda&=&\frac{\sqrt{(r_{\rm ps}^{2}-Q^{2})(r_{\rm ps}^{2}-2Q^{2})}}{\sqrt{3}r_{\rm ps}^{3}}.
\end{eqnarray}
In Fig.~\ref{ppTlambda}, we plot the temperature $T$ as a function of $\Omega$ and $\lambda$, respectively. From Figs. \ref{TOmega} and \ref{Tlambda}, we find that the behavior of the temperature is similar. At first it increases with $\Omega$ or $\lambda$. After its maximum is approached, it will decrease. So there are two different black hole phases bounded by the maximum of the temperature.  At the maximum of the temperature, the parameters have values
\begin{eqnarray}
 T_{\rm M}=\frac{1}{6\sqrt{3}\pi Q},\quad
 \Omega_{\rm M}=\frac{1}{2Q}\sqrt{\frac{2\sqrt{3}}{3}-1},\quad
 \lambda_{\rm M}=\frac{1}{2Q}\sqrt{3-\frac{5}{\sqrt{3}}}.
\end{eqnarray}
Correspondingly, the radius of the photon sphere for this case is
\begin{eqnarray}
 r_{\rm psM}=(1+\sqrt{3})Q.
\end{eqnarray}
Finally, with the help of (\ref{rpsq}), we get the point corresponding to the maximum of the temperature
\begin{eqnarray}
 Q_{\rm M}=\frac{\sqrt{3}}{2}M.
\end{eqnarray}
Obviously, this point exactly matches the Davies point (\ref{Dav}), where the heat capacity diverges.

In summary, we find a new relation between the black hole thermodynamics and dynamics. The phase transition point or Davies point is exactly located at the maximum in the $T$-$\Omega$ or $T$-$\lambda$ plane. This may provide us a new dynamic way to precisely probe the Davies point.

\section{Higher dimensional charged Reissner-Nordstr\"{o}m black holes}
\label{higherdime}

For the $d$-dimensional charged black hole, the line element is just in the same form of (\ref{metric}) while with the metric function given by
\begin{eqnarray}
 f(r)=1-\frac{m}{r^{d-3}}+\frac{q^{2}}{r^{2(d-3)}}.\label{RN}
\end{eqnarray}
The parameters $m$ and $q$ are linked to the black hole mass $M$ and charge $Q$ as
\begin{eqnarray}
 m&=&\frac{16\pi M}{(d-2)A_{d-2}}, \quad  \\
 q&=&\frac{8\pi Q}{\sqrt{2(d-2)(d-3)A_{d-2}}},
\end{eqnarray}
where $A_{d-2}=2\pi^{(d-1)/2}/\Gamma\left((d-1)/2\right)$ is the area of the unit ($d$-2) sphere. The horizons located at the roots of $f(r)=0$
\begin{eqnarray}
 r_{\pm}^{d-3}=\frac{1}{2}\left(m\pm\sqrt{m^{2}-4q^{2}}\right).
\end{eqnarray}
Similar to the four-dimensional case, the higher dimensional black hole has two horizons for $q/m<\frac{1}{2}$, one horizon for $q/m=\frac{1}{2}$, or no horizon for $q/m>\frac{1}{2}$. The temperature and entropy corresponding to the outer horizon are
\begin{eqnarray}
 T&=&\frac{(d-3)(r_{+}^{2d}-q^{2}r_{+}^{6})}{4\pi r_{+}^{2d+1}},\\
 S&=&\frac{A_{d-2}r_{+}^{d-2}}{4}.
\end{eqnarray}
The heat capacity at fixed charge $Q$ is
\begin{eqnarray}
 C_{Q}=\frac{A_{d-2}(d-2)(r_{+}^{2d}-q^{2}r_{+}^{6})}
     {4(2d-5)q^{2}r_{+}^{8-d}-4 r_+^{d+2}}.
\end{eqnarray}
The behavior of this heat capapcity is similar to the four-dimensional black hole case. For small charge, $C_{Q}$ is negative, while for large charge, $C_{Q}$ becomes positive. Moreover, $C_{Q}$ diverges at
\begin{eqnarray}
 q^{2}=\frac{r_{+}^{2(d-3)}}{2d-5}.
\end{eqnarray}
We list the Davies point in Table~\ref{tab1} for $d$=5-10.

\begin{figure}
\center{\subfigure[$d=5$]{\label{omelamba5}
\includegraphics[width=5cm]{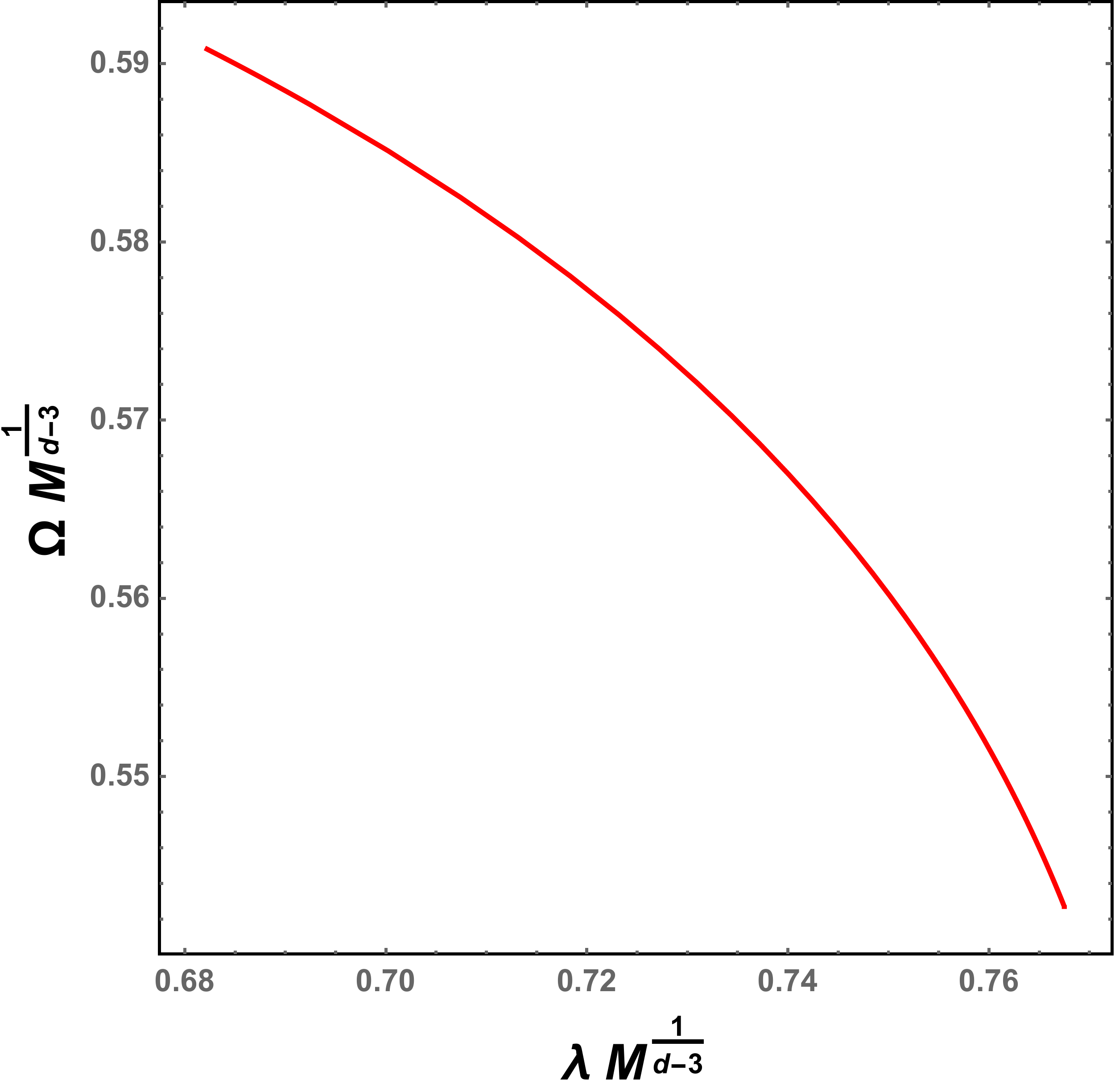}}
\subfigure[$d=6$]{\label{omelambb6}
\includegraphics[width=5cm]{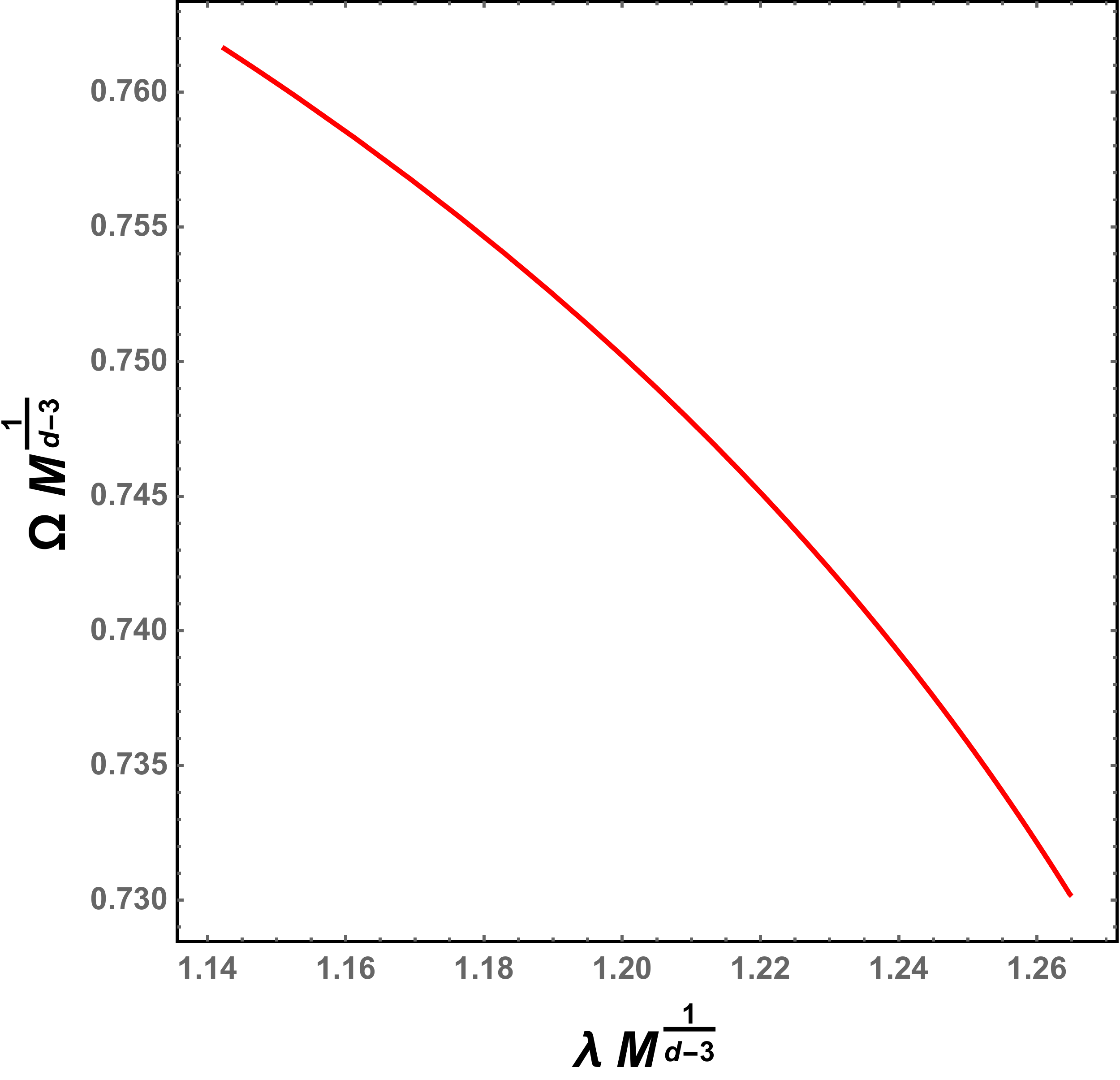}}
\subfigure[$d=7$]{\label{omelambc7}
\includegraphics[width=5cm]{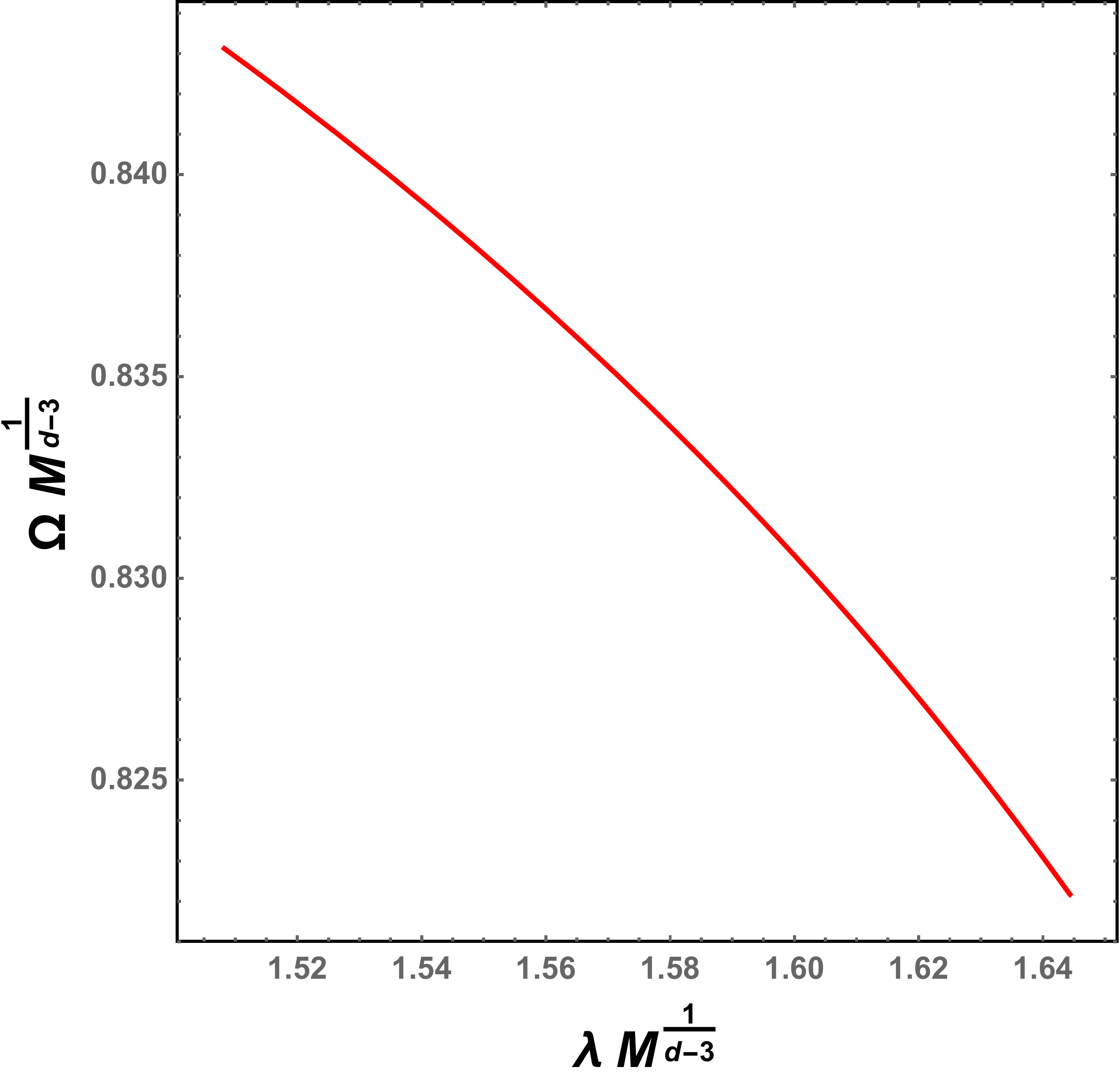}}\\
\subfigure[$d=8$]{\label{omelambd8}
\includegraphics[width=5cm]{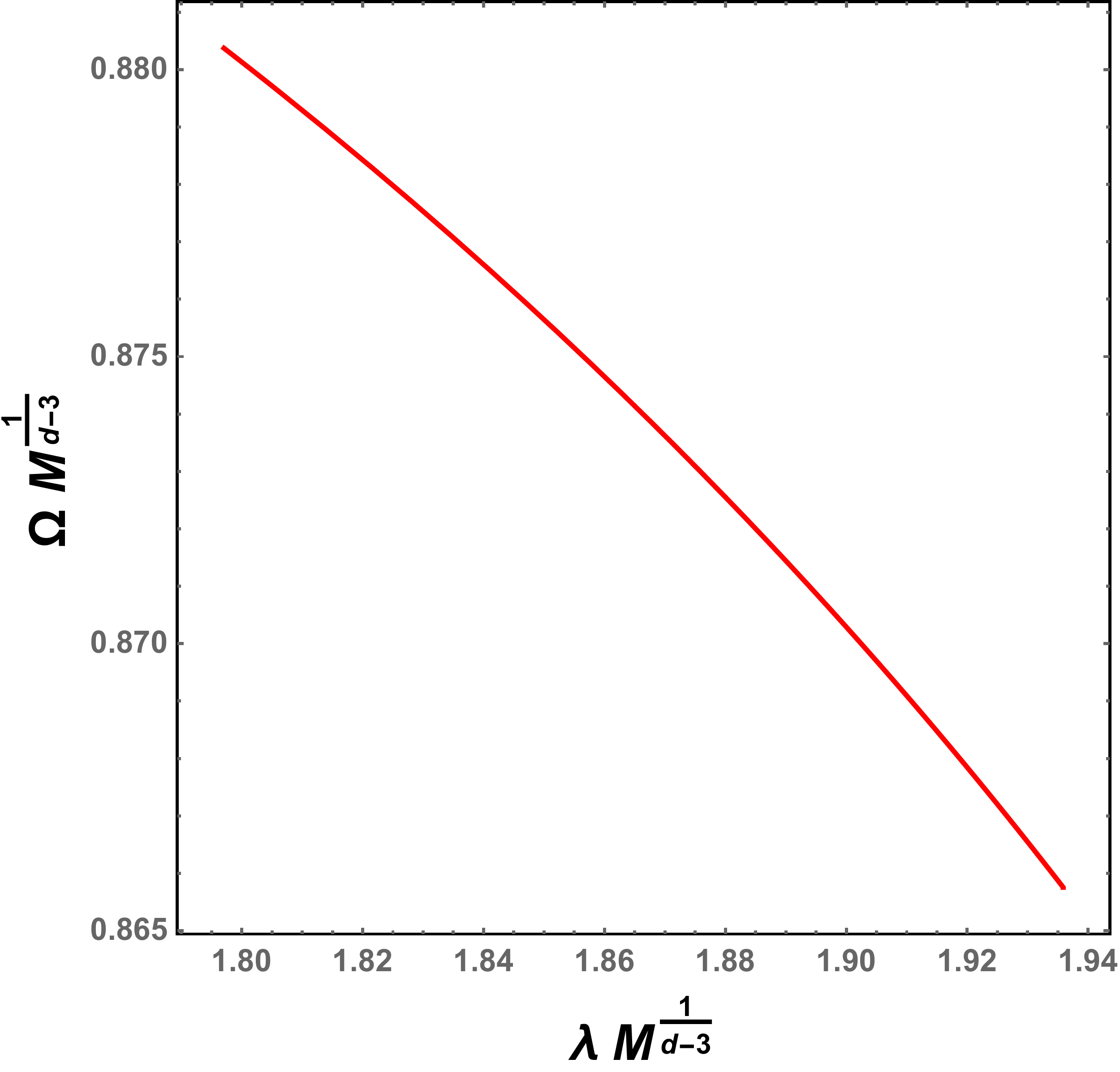}}
\subfigure[$d=9$]{\label{omelambe9}
\includegraphics[width=5cm]{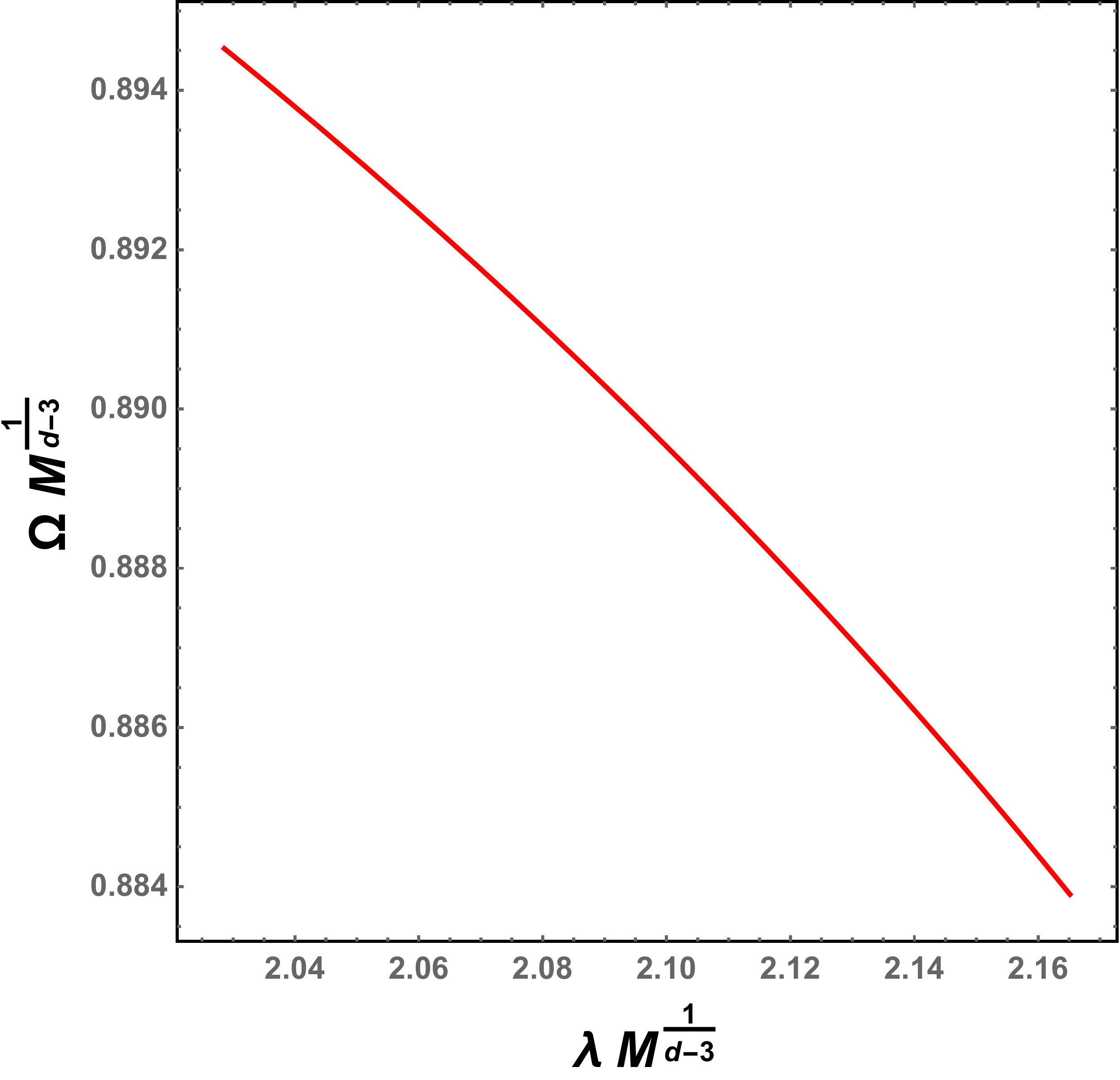}}
\subfigure[$d=10$]{\label{omelambf10}
\includegraphics[width=5cm]{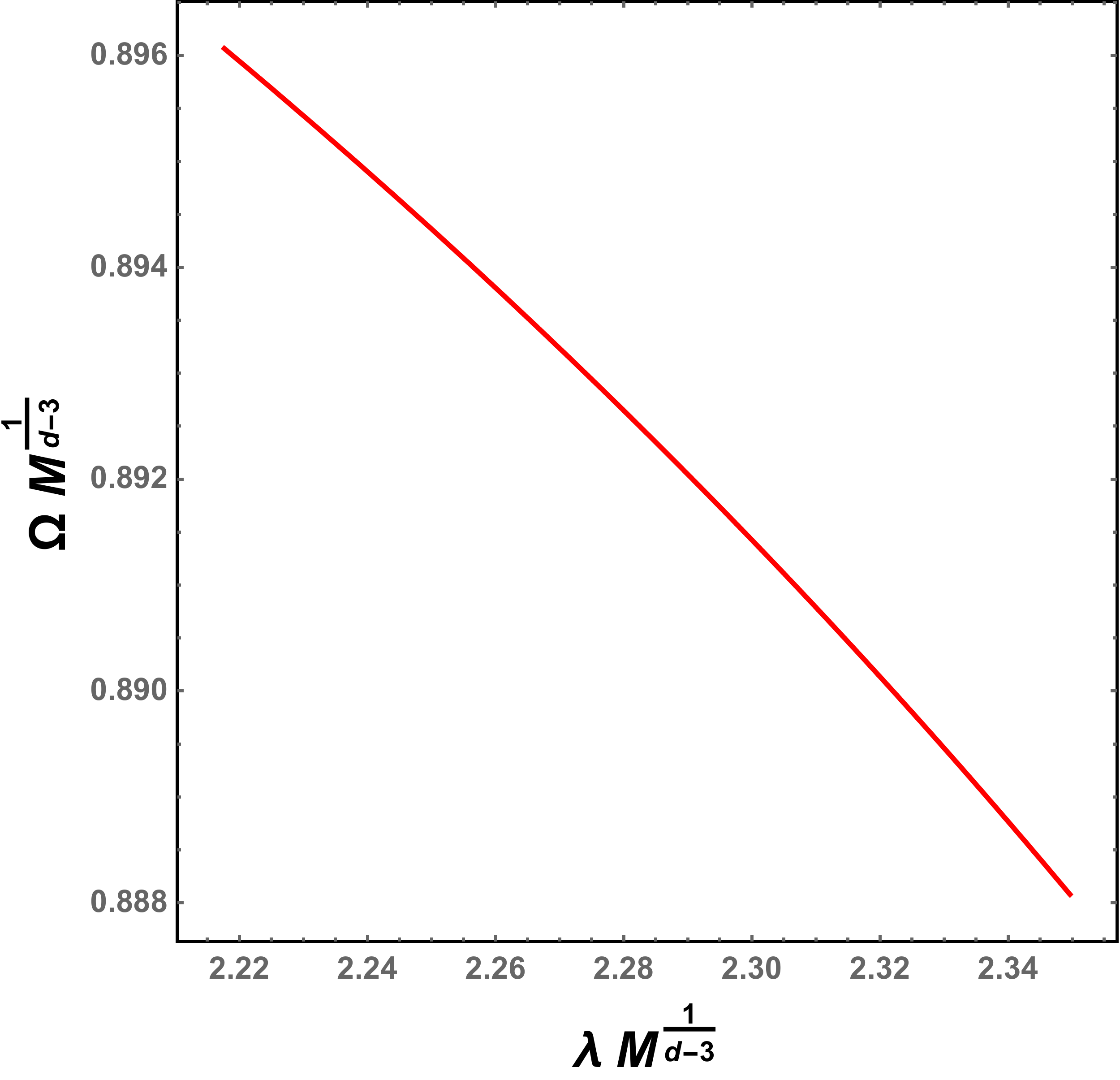}}}
\caption{The angular velocity $\Omega$ and Lyapunov exponent $\lambda$ in the $\Omega$-$\lambda$ plane. The black hole charge $Q$ increases from bottom right to top left. The maximum bound of the charge is 1, $\frac{2}{\sqrt{3}}$, $\sqrt{\frac{3}{2}}$, $2\sqrt{\frac{2}{5}}$, $\sqrt{\frac{5}{3}}$, $2\sqrt{\frac{3 }{7}}$, and $\frac{\sqrt{7}}{2}$ for $d=$5-10, respectively. (a) $d$=5. (b) $d$=6. (c) $d$=7. (d) $d$=8. (e) $d$=9. (f) $d$=10.}\label{ppomelambf10}
\end{figure}

\begin{figure}
\center{\subfigure[]{\label{TVelocityQ}
\includegraphics[width=6cm]{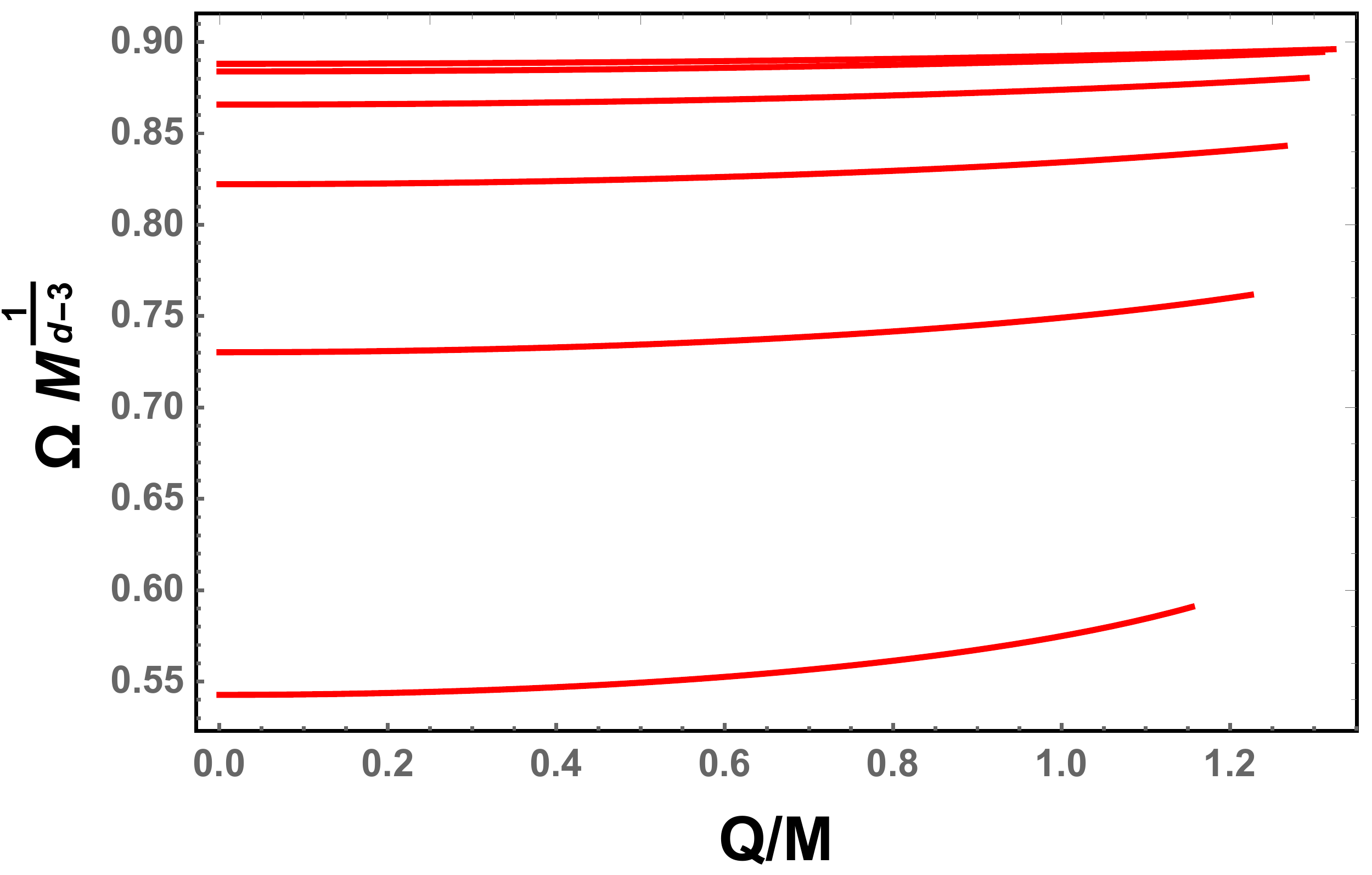}}
\subfigure[]{\label{TLambdaQ}
\includegraphics[width=6cm]{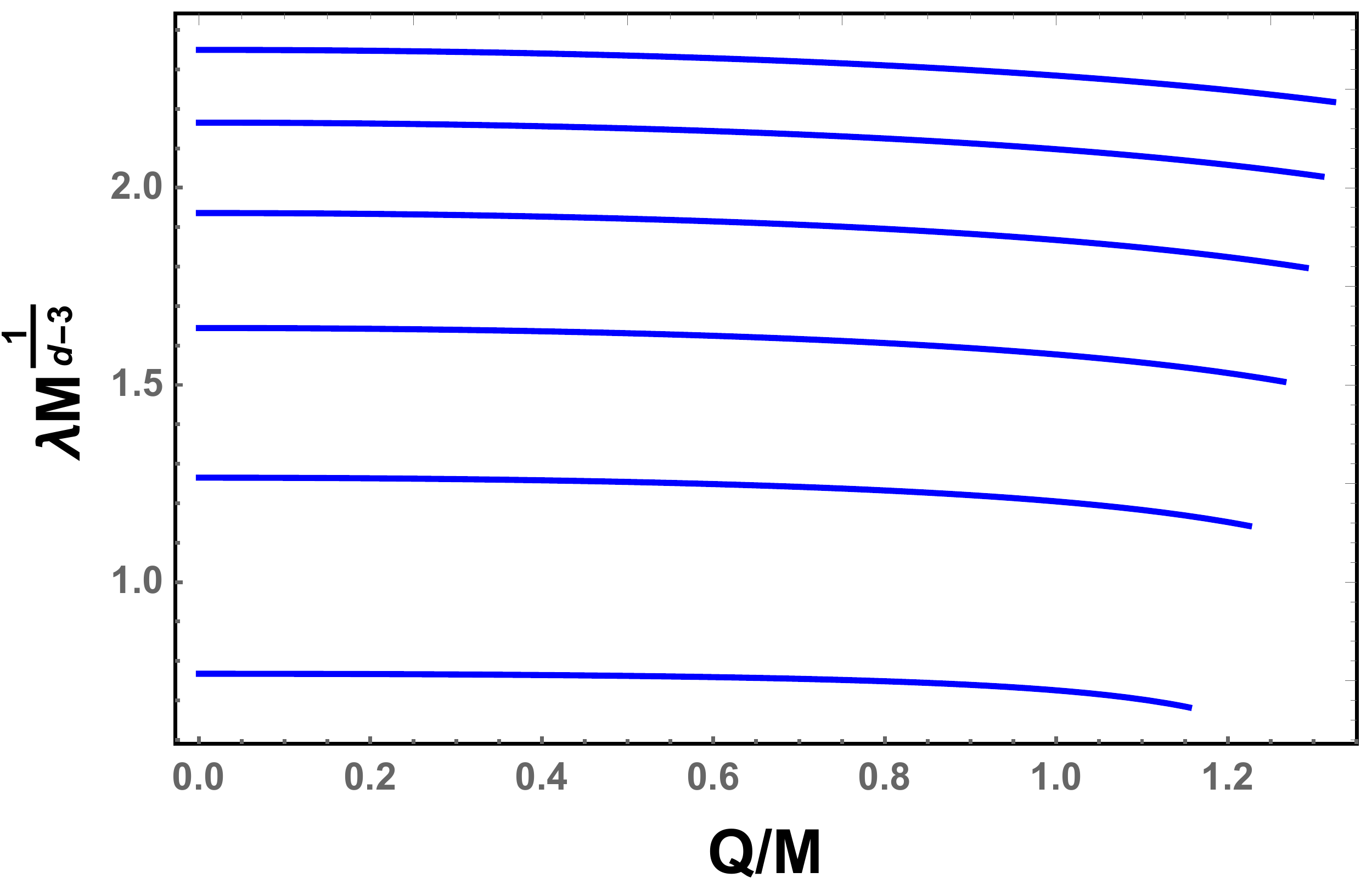}}}
\caption{Behaviors of the angular velocity and Lyapunov exponent as functions of the black hole charge. Spacetime dimension $d$=5-10 from bottom to top. (a) $\Omega$ vs. $Q$. (b) $\lambda$ vs. $Q$.}\label{ppLambdaQ}
\end{figure}

Solving (\ref{rps1}), we can get the radius of the photon sphere
\begin{eqnarray}
 r_{\rm ps}^{d-3}=
 \frac{4\pi}{(d-2)A_{d-2}}\left((d-1)M+
 \sqrt{(d-1)^{2}M^{2}-\frac{2(d-2)^{2}Q^{2}}{d-3}}\right).
\end{eqnarray}
Moreover, the angular velocity and Lyapunov exponent are
\begin{eqnarray}
 \Omega^{2}&=&\frac{1+q^{2}r_{\rm ps}^{6-2d}-mr_{\rm ps}^{3-d}}{r_{\rm ps}^{2}},\\
 \lambda^{2}&=&\frac{(q^{2}r_{\rm ps}^{6}+r_{\rm ps}^{2d}-mr_{\rm ps}^{d+3})(2r_{\rm ps}^{2d}
    -2(d-2)(2d-7)q^{2}r_{\rm ps}^{6}+(d-1)(d-4)mr_{\rm ps}^{d+3})}{2r_{\rm ps}^{2(2d+1)}}.
\end{eqnarray}
In Fig.~\ref{ppomelambf10}, we show the angular velocity and Lyapunov exponent in the $\Omega$-$\lambda$ plane for the spacetime dimension $d$=5-10, respectively. Different from the $d$=4 case, there are no the spiral-like shapes for higher dimensional black hole cases, and the angular velocity $\Omega$ is just a monotone decreasing function of $\lambda$. In order to further understand this result, the behaviors of the angular velocity and Lyapunov exponent are presented  in Fig.~\ref{ppLambdaQ} as functions of the black hole charge. Obviously, $\Omega$ increases while $\lambda$ decreases with the charge $Q$. Both them are monotonic functions of the charge Q. So the spiral-like shape does not exist for the higher dimensional charged black holes. On the other hand, we note that in Ref.~\cite{He}, the authors found that, for the five-dimensional charged Kaluza-Klein black hole with squashed horizons, the QNMs demonstrate the spiral-like shapes near the Davies point, which may be caused by the non-vanishing Kaluza-Klein parameters.

\begin{figure}
\center{\subfigure[]{\label{TQ}
\includegraphics[width=6cm]{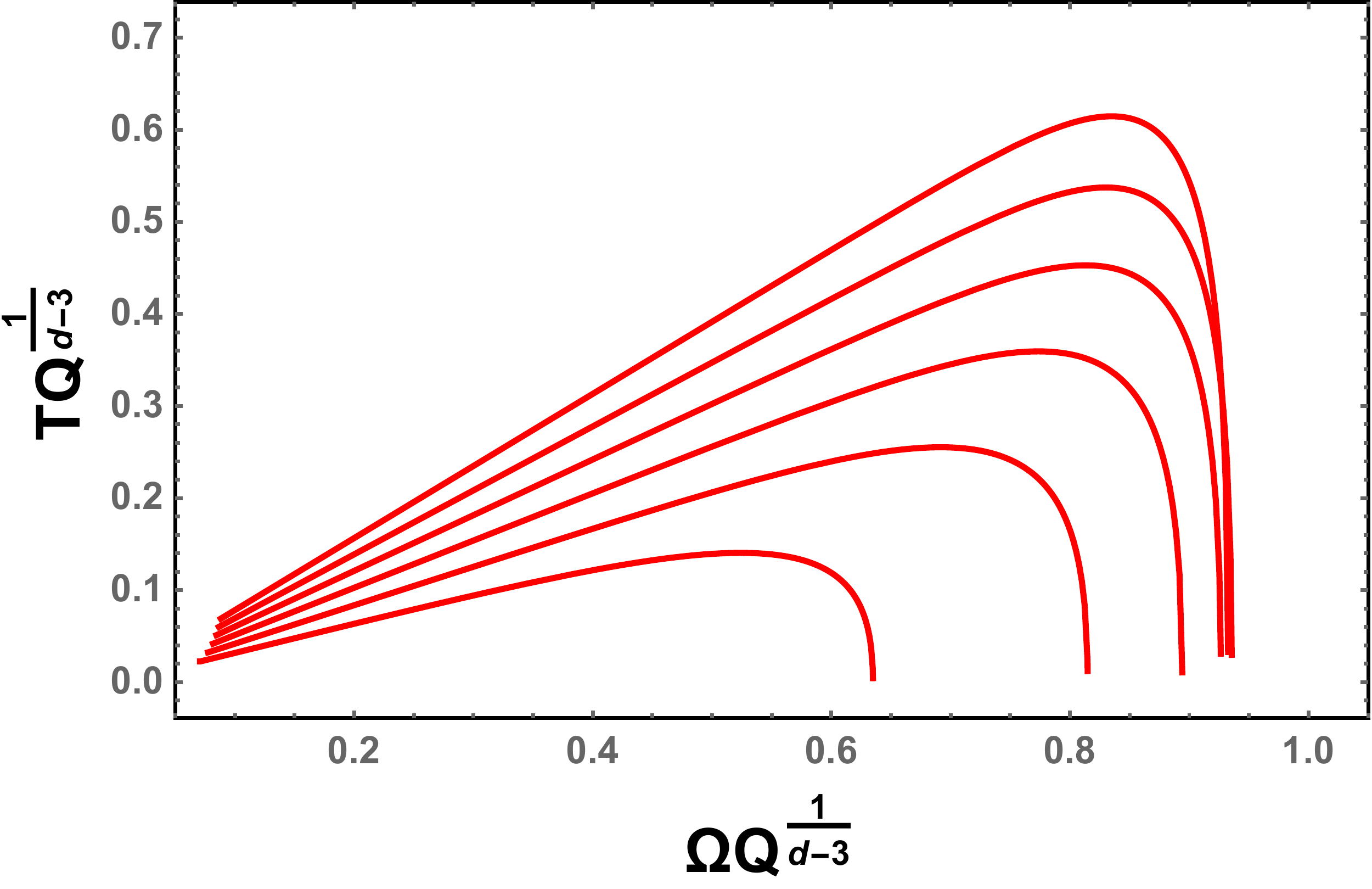}}
\subfigure[]{\label{TLambda}
\includegraphics[width=6cm]{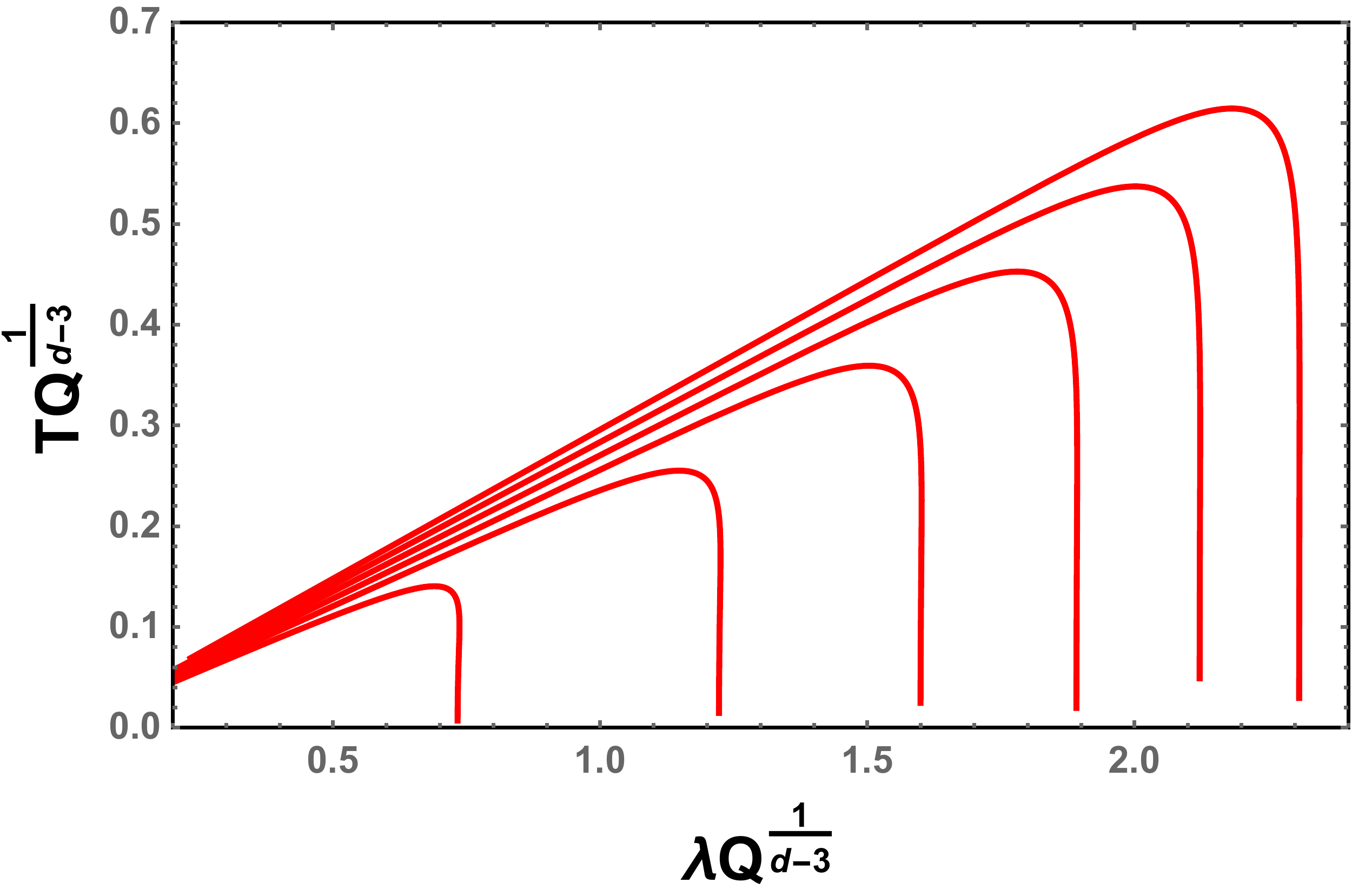}}}
\caption{(a) $T$ vs. $\Omega$. (b) $T$ vs. $\lambda$. Spacetime dimension $d$=5-10 from bottom to top. }\label{ppTLambda}
\end{figure}

In the above section, we present a new relation between the black hole thermodynamics and dynamics for the four-dimensional spacetime. The temperature has a maximum value in the $T$-$\Omega$ or $T$-$\lambda$ plane, and the maximum point is found to be corresponded to the Davies point. Here we wonder whether this property holds for higher dimensional spacetime. To answer this question, we describe the temperature $T$ as a function of $\Omega$ and $\lambda$, respectively, in Fig.~\ref{ppTLambda}. Interestingly, similar to the four-dimensional case, there also exhibit maximum values for the temperature in both the figures for $d$=5-10. Using this maximum values of the temperature, we can obtain the corresponding charge $Q_{\rm M}$. We list the parameter values corresponding to the maximum values of the temperature in Table~\ref{tab1} for $d$=5-10. From the table, we can clearly see that the Davies points and the $Q_{\rm M}$ exactly match each other for $d$=5-10. Thus we confirm that the relation we propose above also holds for the higher dimensional black holes.

\begin{table}[h]
\begin{center}
\begin{tabular}{ccccccc}
  \hline\hline
                & $d$=5 & $d$=6 & $d$=7 & $d$=8 & $d$=9 & $d$=10 \\\hline
 $Q_{\rm D}/M$ & 0.8607 & 0.8101 & 0.7589 & 0.7136 & 0.6744 & 0.6404 \\
 $Q_{\rm M}$/M & 0.8607 & 0.8101 & 0.7589 & 0.7136 & 0.6744 & 0.6404 \\
 $r_{\rm psM}/Q^{\frac{1}{d-3}}$
                   & 1.3189 & 1.1085 & 1.0497 & 1.0361 & 1.0410 & 1.0547 \\
 $T_{\rm M}Q^{\frac{1}{d-3}}$
                   & 0.1404 & 0.2552 & 0.3593 & 0.4528 & 0.5373  & 0.6146 \\
 $\Omega_{\rm M}Q^{\frac{1}{d-3}}$
                   & 0.5240 & 0.6916 & 0.7735 & 0.8130 & 0.8301  & 0.8349 \\
 $\lambda_{\rm M}Q^{\frac{1}{d-3}}$
                   & 0.6897 & 1.1478 & 1.5033 & 1.7806 & 2.0017  & 2.1820 \\ \hline\hline
\end{tabular}
\caption{The Davies points and parameter values corresponding to the maximum value of the temperature $T_{\rm M}$ for higher dimensional charged RN black holes.}\label{tab1}
\end{center}
\end{table}

\section{Charged Reissner-Nordstr\"{o}m dS black holes}
\label{RNdsBH}

Recently, there is a great interest focusing on the strong cosmic censorship in dS spacetime~\cite{Cardosob,Hod,Dias,Cardosoc,Mo,Diasb,Luna,Gej,Destounis,LiuTang}. Many works showed that this censorship may be violated for the charged RN-dS black hole by calculating the QNMs of different types of perturbations. In particular, in Ref.~\cite{Cardosob}, the authors divided the QNMs for the charged RN-dS black hole into three families: the photon sphere modes, dS modes, and near-extremal modes. Therefore, it is very interesting to examine the relation presented in above section by calculating the angular velocity and the Lyapunov exponent of the photon sphere modes.

Here we only consider the four-dimensional spacetime. The charged RN-dS black hole solution is also described by the metric (\ref{metric}) while with the function
\begin{eqnarray}
 f(r)=1-\frac{2M}{r}+\frac{Q^{2}}{r^{2}}-\frac{\Lambda}{3}r^{2}.
\end{eqnarray}
Here, the parameter $\Lambda$ is the cosmological constant, which is positive for the dS spacetime. For this case, there exists a cosmological horizon outer the event horizon, and both them can be obtained by solving $f(r)=0$. Moreover, we can express the mass $M$ with the radius $r_{+}$ of the event horizon
\begin{eqnarray}
 M=\frac{3Q^{2}+3r_{+}^{2}-r_{+}^{4}\Lambda}{6r_{+}}.
\end{eqnarray}
The black hole entropy corresponding to the event horizon is still a quarter of its area $S=A/4=\pi r_{+}^{2}$.  Thus, the mass can be further expressed as
\begin{eqnarray}
 M=\frac{3\pi^{2} Q^{2}+3\pi S-S^{2}\Lambda}{6\pi^{3/2}S^{1/2}}.\label{mm}
\end{eqnarray}
Similarly, the temperature of the event horizon is
\begin{eqnarray}
 T=\frac{-\pi^{2}Q^{2}+\pi S-S^{2}\Lambda}{4(\pi S)^{3/2}}.
\end{eqnarray}
The heat capacity at fixed charge $Q$ and cosmological $\Lambda$ is
\begin{eqnarray}
 C_{Q,\Lambda}
          =\frac{2S(\pi^{2}Q^{2}-\pi S+S^{2}\Lambda)}{-3\pi^{2}Q^{2}+\pi S+S^{2}\Lambda}.
\end{eqnarray}
Note that $C_{Q,\Lambda}$ exactly vanishes at $T$=0, which corresponds to the extremal black hole case. It diverges at the point where its denominator vanishes, i.e.,
\begin{eqnarray}
 S=\frac{\pi(\sqrt{1+12Q^{2}\Lambda}-1)}{2\Lambda}.
\end{eqnarray}
This is just the Davies point. Plugging it into Eq.~(\ref{mm}), we can find another form of the Davies point
\begin{eqnarray}
 \Lambda M^{2}=\frac{2}{9}\left(\sqrt{12Q^{2}\Lambda+1}-1\right),
\end{eqnarray}
or
\begin{eqnarray}
 \left(\frac{Q_{\rm D}}{M}\right)^{2}=\frac{3}{4}+\frac{27}{16}\Lambda M^{2}.\label{DDavies}
\end{eqnarray}
Obviously, when the cosmological constant $\Lambda\rightarrow0$, this result will reduce back to that of the RN black hole case (\ref{Dav}). It is worthwhile noting that this Davies point (\ref{DDavies}) is different from that given in~\cite{Daviespcc}, where the coefficient of the second term in the right side of the corresponding equation is $\frac{81}{16}$ but not $\frac{27}{16}$. The reason is that Davies made a change $\Lambda\rightarrow 3\Lambda$. So both results are consistent with each other.

Now let us turn to the null geodesics for the charged dS black hole. By solving (\ref{rps1}), we obtain the radius of the photon sphere,
\begin{eqnarray}
 r_{\rm ps}=\frac{1}{2}\left(3M+\sqrt{9M^{2}-8Q^{2}}\right).\label{rppss}
\end{eqnarray}
Interestingly, this result is just the form (\ref{rpsq}) for the asymptotically flat case without the cosmological constant. The reason of this is not hard to understand. From (\ref{veff}), we can find that the cosmological constant $\Lambda$ term only comes to the effective potential as a constant. The photon sphere is obtained by solving $\partial_{r}V_{\rm eff}=0$, so after the derivation, the effect of $\Lambda$ in the metric function $f(r)$ disappears, which leads to the same form of the photon sphere as the asymptotically flat case. Nevertheless, one needs keep in mind that the mass $M$ here indeed depends on the cosmological constant, see Eq. (\ref{mm}).

\begin{figure}
\center{\includegraphics[width=6cm]{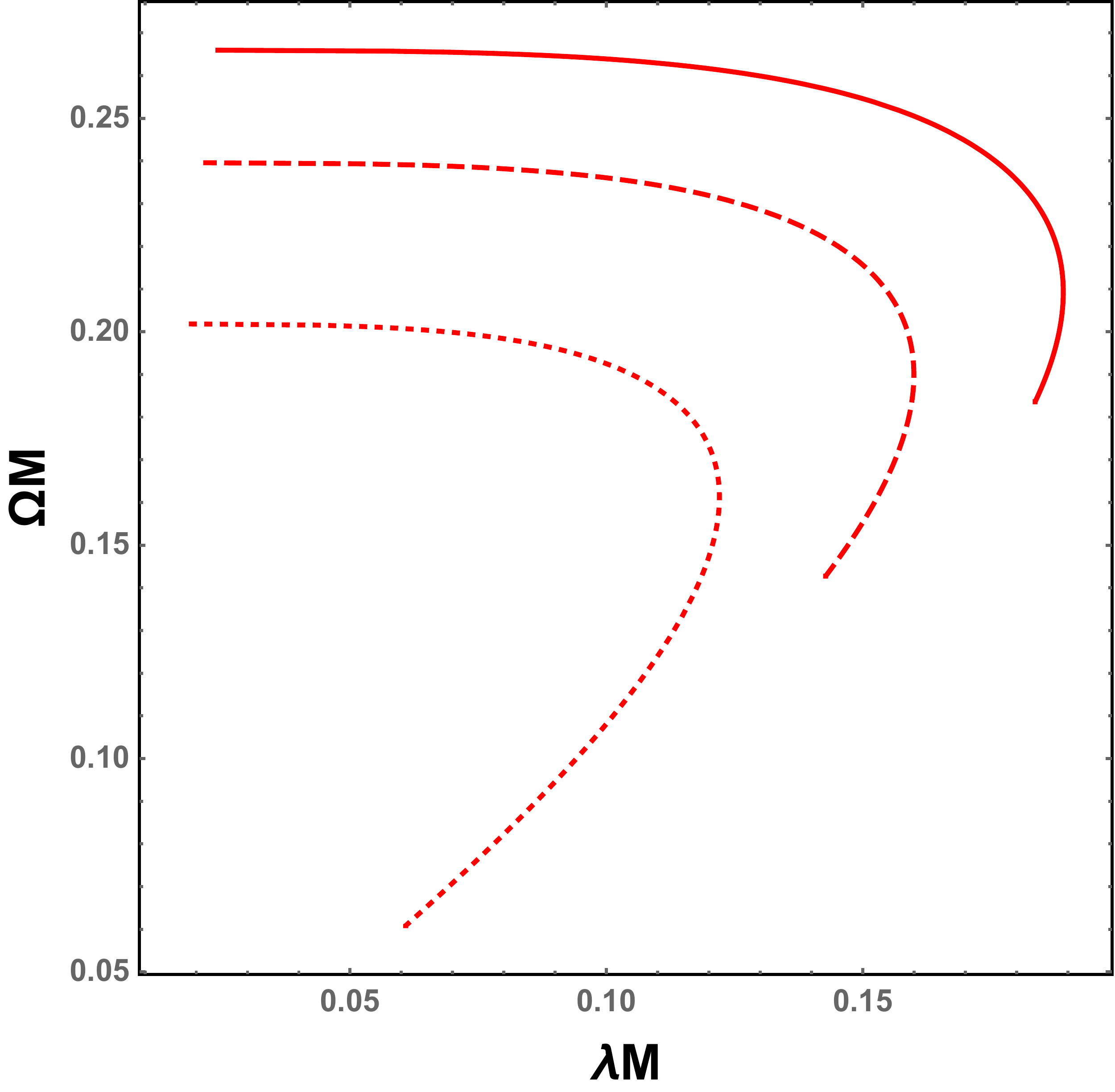}}
\caption{The angular velocity and Lyapunov exponent in the $\Omega$-$\lambda$ plane for $\Lambda M^{2}$=0.01 (solid line), 0.05 (dashed line), and 0.1 (dotted line) for the charged RN-dS black hole.}\label{OmeLamds}
\end{figure}

\begin{figure}
\center{\subfigure[]{\label{TOQds}
\includegraphics[width=6cm]{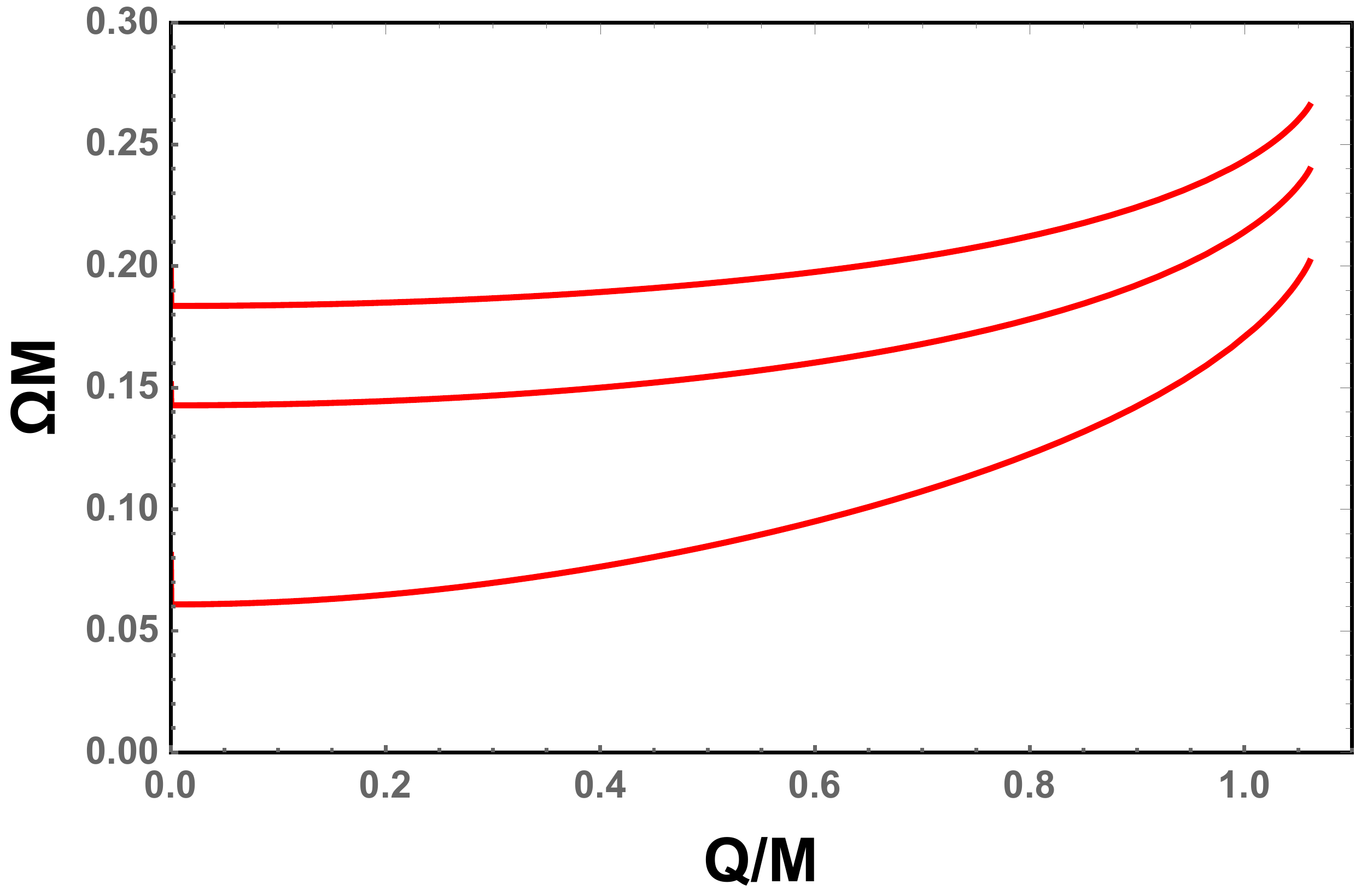}}
\subfigure[]{\label{TLQds}
\includegraphics[width=6cm]{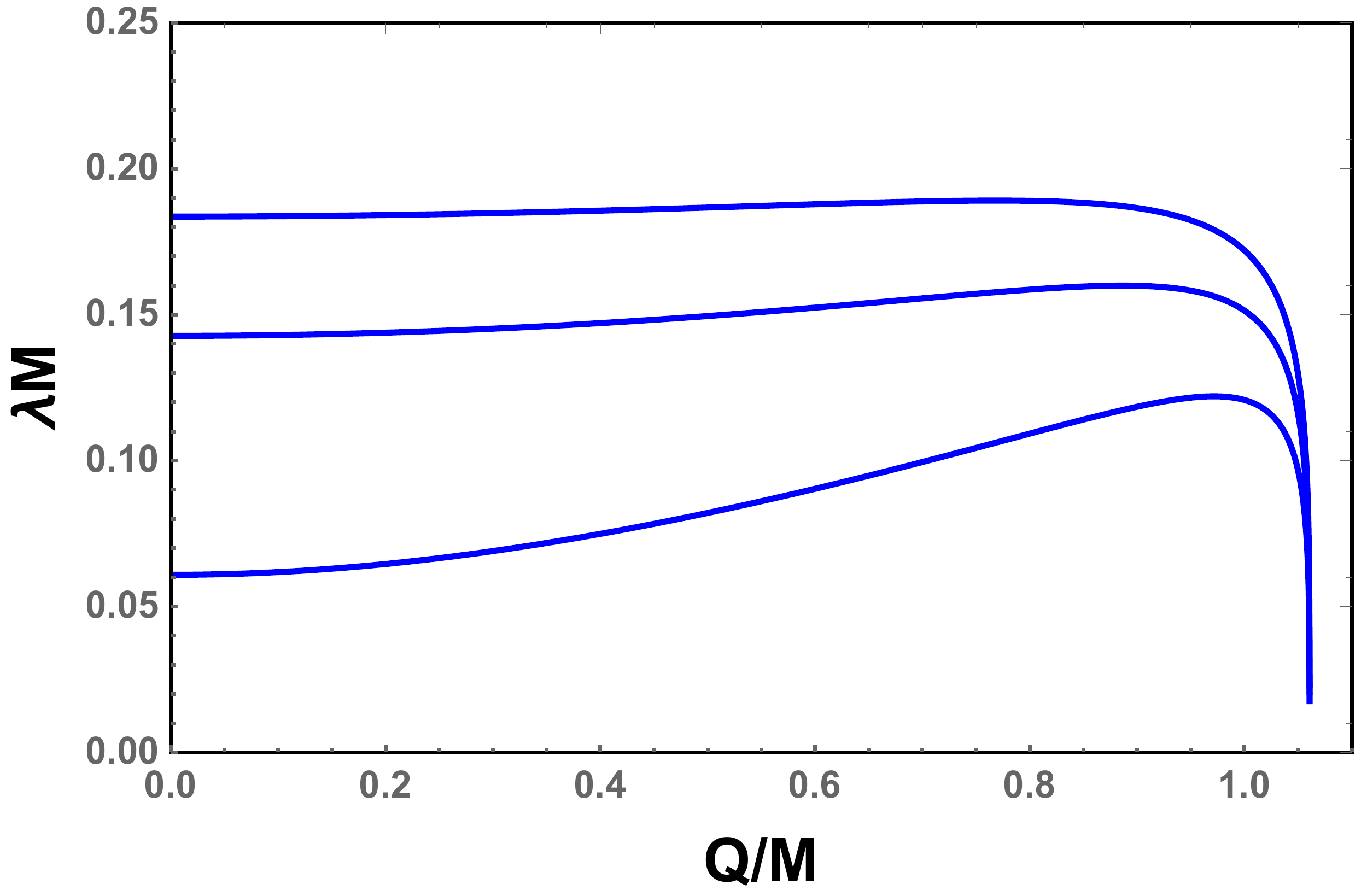}}}
\caption{Behaviors of the angular velocity and Lyapunov exponent as functions of the black hole charge for $\Lambda M^{2}$=0.01, 0.05, and 0.1, respectively, from top to bottom.}\label{ppOQds}
\end{figure}

Adopting the form (\ref{rppss}), we can further obtain the angular velocity and the Lyapunov exponent,
\begin{eqnarray}
 \Omega&=&\frac{\sqrt{-9 M^2 \left(4 \Lambda
   Q^2+1\right)+3 M \sqrt{9 M^2-8 Q^2}
   \left(1-4 \Lambda  Q^2\right)+4 Q^2 \left(4
   \Lambda  Q^2+3\right)}}{\sqrt{6} Q
   \left(\sqrt{9 M^2-8 Q^2}+3 M\right)},\\
 \lambda&=&\frac{4 \sqrt{8 Q^2-3 M \left(\sqrt{9
   M^2-8 Q^2}+3 M\right)}}{\sqrt{3}
   \left(\sqrt{9 M^2-8 Q^2}+3 M\right)^3}\nonumber\\
   &&\times
   \sqrt{ 81
   \Lambda  M^4-9 M^2 \left(8 \Lambda
   Q^2+1\right)+3 M \sqrt{9 M^2-8 Q^2} \left(9
   \Lambda  M^2-4 \Lambda  Q^2-1\right)+8
   \Lambda  Q^4+6 Q^2}.
\end{eqnarray}
We plot the behavior of the angular velocity and Lyapunov exponent in the $\Omega$-$\lambda$ plane for $\Lambda M^{2}$=0.01, 0.05, and 0.1 in Fig.~\ref{OmeLamds}. For fixed $\Lambda M^{2}$, it is obvious that there exist spiral-like shapes in this plane, which is similar to the asymptotically flat case. The existence of the spiral-like shapes is also the nonmonotonic behavior of $\lambda$ rather $\Omega$. For clarity, we show them in Fig.~\ref{ppOQds}. The values of the charge at the starting point of the spiral-like shapes for $\Lambda M^{2}$=0.01, 0.05, and 0.1 are $Q_{\rm S}/M$=0.7684, 0.8862, and 0.9716, which are different from the Davies points $Q_{\rm D}/M$=0.8757, 0.9134, and 0.9585. Therefore, the result clearly shows that the starting points of the spiral-like shapes and the Davies points do not match each other.

\begin{figure}
\center{\subfigure[]{\label{TTome}
\includegraphics[width=6cm]{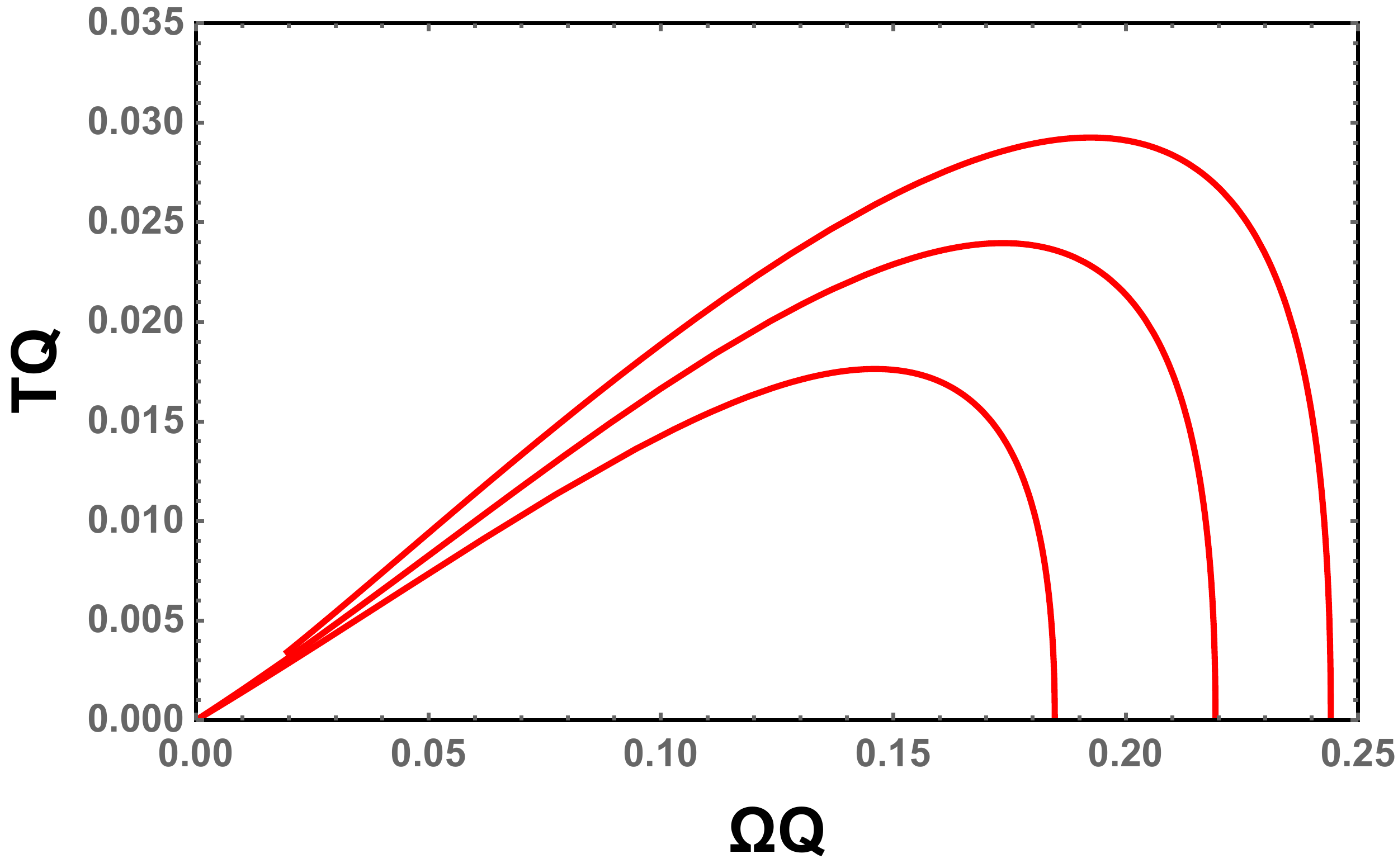}}
\subfigure[]{\label{TTLam_10b}
\includegraphics[width=6cm]{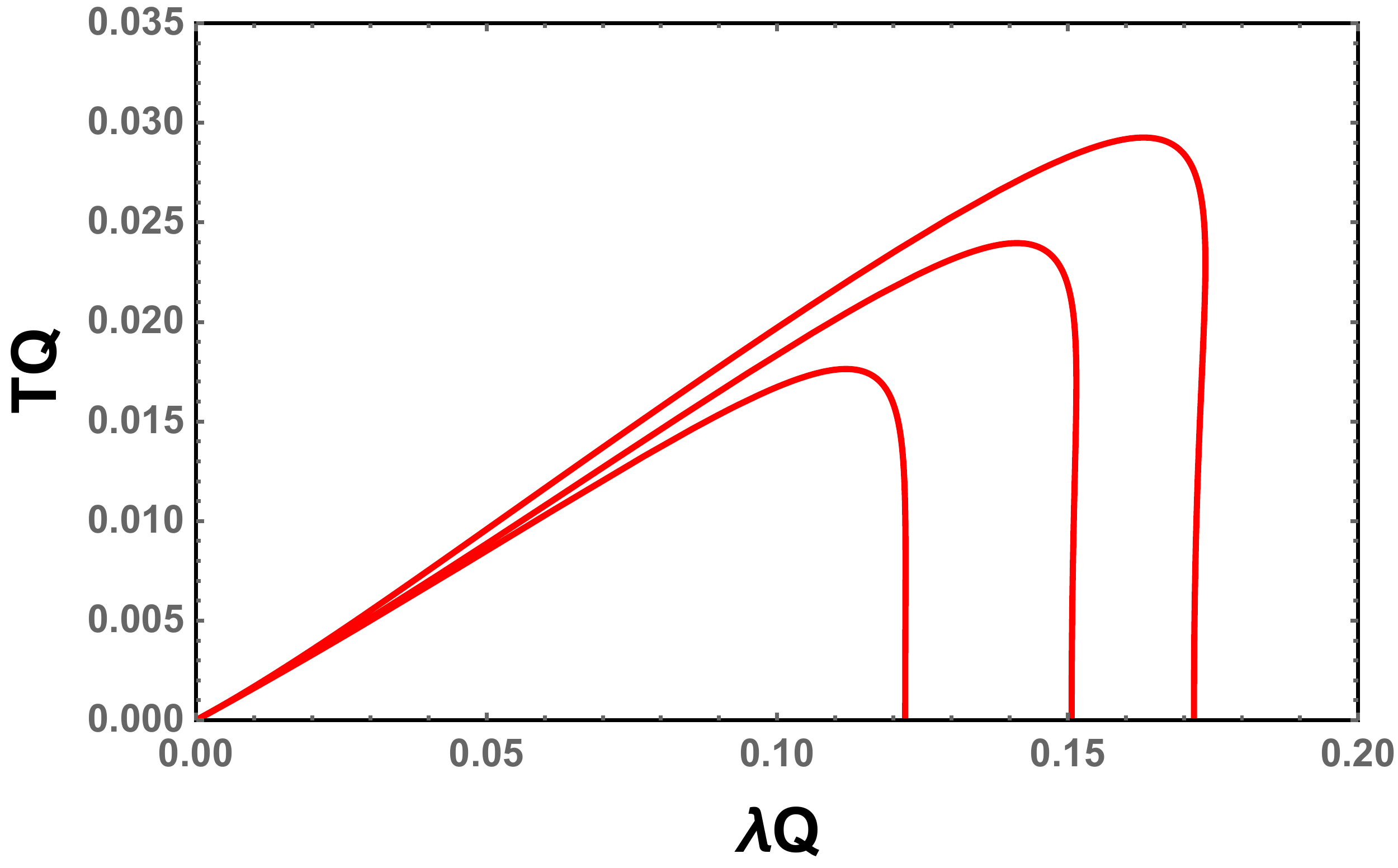}}}
\caption{Behaviors of the temperature $T$ as a function of $\Omega$ and $\lambda$ for $\Lambda Q^{2}$=0.01, 0.05, and 0.1 from top to bottom. (a) $T$-$\Omega$. (b) $T$-$\lambda$}\label{ppTome}
\end{figure}

In Fig.~\ref{ppTome}, we describe the behavior of the temperature as a function of $\Omega$ and $\lambda$, respectively. From it, there is a local maximum for each fixed $\Lambda Q^{2}$. With the increase of $\Lambda Q^{2}$, the maximum decreases and is shifted to small value of $\Omega$ or $\lambda$. Moreover, the values for different parameters corresponding to the maximum of the temperature are listed in Table~\ref{tab2}. It is clear that the Davies point and starting point of the spiral-like shape are not consistent with each other. For small $\Lambda$, $Q_{\rm S}$ is smaller than $Q_{\rm D}$. However, with the increase of $\Lambda$, $Q_{\rm S}$ increases faster than $Q_{\rm D}$. For example, for $\Lambda Q^{2}=0.1$, $Q_{\rm S}$ will be larger than $Q_{\rm D}$. On the other hand, the value of the charge $Q_{\rm M}$ corresponding to the maximum exactly coincides with the Davies point $Q_{\rm D}$ given in (\ref{DDavies}). Therefore, this further confirms our new relation between the black hole thermodynamics and dynamics shown in Sec. \ref{newrelation}. In summary, this conjecture is not only effective for the asymptotically flat spacetime, but also for the asymptotically dS spacetime.

\begin{table}[h]
\begin{center}
\begin{tabular}{ccccccc}
  \hline\hline
                & $\Lambda Q^{2}$=0.01 & $\Lambda Q^{2}$=0.05 & $\Lambda Q^{2}$=0.1  \\\hline
 $Q_{\rm D}/M$ & 0.8786 & 0.9216 & 0.9650 \\
 $Q_{\rm S}$/M & 0.7906 & 0.9085 & 0.9774 \\
 $Q_{\rm M}$/M & 0.8786 & 0.9216 & 0.9650 \\\hline
 $r_{\rm psM}/Q$
                   & 2.6639 & 2.4333 & 2.1996 \\
 $T_{\rm M}Q$
                   & 0.0293 & 0.0240 & 0.0176 \\
 $\Omega_{\rm M}Q$
                   & 0.1924 & 0.1736 & 0.1460 \\
 $\lambda_{\rm M}Q$
                   & 0.1631 & 0.1412 & 0.1118 \\ \hline\hline
\end{tabular}
\caption{The charge of the Davies point and starting point of the spiral-like shape of the charged RN-dS black holes. Other parameter values correspond to the maximum value of the temperature $T_{\rm M}$.}\label{tab2}
\end{center}
\end{table}

\section{Conclusions and discussions}
\label{Conclusion}

In this paper, we devoted to explore the relation between the spiral-like shape of the QNMs and Davies point, where a phase transition takes place between a thermodynamic stable phase and an unstable phase.

First, we gave a brief view on the link between the QNMs and the photon sphere. In the eikonal limit, this link is thought to be very accurate. This gives us a possible way to obtain the QNMs for the black hole from the side of the photon sphere. Then the angular velocity and the Lyapunov exponent, respectively, corresponding to the real part and imaginary part of the QNMs, are parameterized by the parameter of the black hole photon sphere. This study is analytic or exact for the charged RN black hole and RN-dS black hole in four or higher dimensional spacetime. So it can exactly confirm the relation that we expect.

For the four-dimensional charged RN black hole, there exhibits a simple spiral-like behavior in the $\Omega$-$\lambda$ plane. The reason for this is because that the Lyapunov exponent $\lambda$ has a maximum as the black hole charge $Q$ changes. While the angular velocity $\Omega$ is just a monotonic increasing function of the black hole charge. This result is very different from the numerical results given in Ref.~\cite{Jing}, where both $\Omega$ and $\lambda$ are non-monotonic functions of the charge, and this will lead to an explicit spiral-like behavior in the $\Omega$-$\lambda$ plane. Nevertheless, the authors claimed that the Davies point is linked to the starting point of this spiral-like shape. Their numerical result states that the difference between these two points is very small indicating a correspondence exists. We also analytically examined this issue in detail. The result shows that the starting point of the spiral-like shape is $Q_{\rm S}=\frac{\sqrt{51-3\sqrt{33}}}{8}M\approx0.7264M$, which has about 16\% deviation from the Davies point $Q_{\rm D}=\frac{\sqrt{3}}{2}\approx0.8660M$. Therefore, the Davies point and the starting point of the spiral-like shape are not exactly coincide. However, we further investigated the behavior of the black hole temperature as a function of the angular velocity and Lyapunov exponent. The temperature in both the figures clearly demonstrates the local maximum. More interestingly, the local maximum analytically coincides with the Davies point. This indicates that there is a relation between the Davies point and the maximum of the temperature in the $T$-$\Omega$ and $T$-$\lambda$ planes. Since both $\Omega$ and $\lambda$ correspond to the QNMs of the black hole, this relation may provide an exact correspondence between the black hole thermodynamics and dynamics.

We also extended our investigation to the higher dimensional spacetime. One interesting result different from the four-dimensional case is that, in the $\Omega$-$\lambda$ plane, there does not exist the spiral-like shape. So the relation between the Davies point and the behavior of the spiral-like shape is completely lost. Fortunately, the relation we proposed between the Davies point and the maximum of the temperature in the $T$-$\Omega$ and $T$-$\lambda$ planes still holds for the higher dimensional charged black holes.

Moreover, we also applied the study to the charged RN black hole in the dS spacetime, and the results are similar to the asymptotically flat spacetime. The Davies point, the starting point of the spiral-like shape, and the maximum of the temperature in the $T$-$\Omega$ and $T$-$\lambda$ planes are all found to increase with the cosmological constant. As expected, the Davies point and the starting point of the spiral-like shape does not coincide with each other. While the Davies point exactly coincides with the maximum of the temperature, which further confirms our new relation even in the dS spacetime.

Before ending this paper, we would like to give a few comments. First, at least in the eikonal limit, there is no exact relation between the Davies point and the spiral-like shape of the QNMs. Second, there is a non-trivial and exact relation between the Davies point and the maximum of the temperature in the $T$-$\Omega$ and $T$-$\lambda$ planes. Moreover, some other relations between the thermodynamics and dynamics are worth to be explored. With the further direct observation of the gravitational waves by LIGO and Virgo collaborations, it may provide us a possible way to test the black hole thermodynamics through these relations.

\section*{Acknowledgements}
We would like to thank Profs. Qiyuan Pan and Robert B. Mann for the useful discussions. This work was supported by the National Natural Science Foundation of China (Grants No. 11675064, No. 11875151, and No. 11522541). S.-W. Wei was also supported by the Chinese Scholarship Council (CSC) Scholarship (201806185016) to visit the University of Waterloo.


\begin{thebibliography}{99}

\bibitem{Abbott}
 B. P. Abbott \emph{et al} (Virgo, LIGO Scientific),
 {\em Observation of gravitational waves from a binary black hole merger},
   Phys. Rev. Lett. \textbf{116}, 061102 (2016), [arXiv:1602.03837 [gr-qc]].

\bibitem{Hawking}
 S. W. Hawking,
 {\em Particle creation by black holes},
   Commun. Math. Phys. \textbf{43}, 199 (1975).

\bibitem{Bekensteina}
 J. Bekenstein,
 {\em Black holes and the second law},
     Lett. Nuovo Cim. \textbf{4}, 737 (1972).

\bibitem{Bekensteinb}
 J. D. Bekenstein,
 {\em Black holes and entropy},
    Phys. Rev. D \textbf{7}, 2333 (1973).

\bibitem{Bardeen}
  J. M. Bardeen, B. Carter, and S. Hawking,
   {\em The four laws of black hole mechanics},
    Commun. Math. Phys. \textbf{31}, 161 (1973).

\bibitem{Daviespc}
  P. C. W. Davies,
 {\em The thermodynamic theory of black holes},
Proc. Roy. Soc. Lond. A \textbf{353}, 499 (1977).

\bibitem{Daviespcb}
  P. C. W. Davies,
  {\em Thermodynamics of black holes},
    Rep. Prog. Phys. \textbf{41}, 1313 (1978).

\bibitem{Daviespcc}
 P. C. W. Davies,
 {\em Thermodynamic phase transitions of Kerr-Newman black holes in de Sitter space},
 Class. Quant. Grav. \textbf{6}, 1909 (1989).

\bibitem{Koutsoumbas}
 G. Koutsoumbas, S. Musiri, E. Papantonopoulos, and G. Siopsis,
 {\em Quasi-normal modes of electromagnetic perturbations of four-dimensional topological black holes with scalar hair},
 J. High Energy Phys. \textbf{0610}, 006 (2006),
 [arXiv:hep-th/0606096].

\bibitem{Koutsoumbasb}
G. Koutsoumbas, E. Papantonopoulos, and G. Siopsis,
   {\em phase transitions in charged topological-AdS black holes},
   J. High Energy Phys. \textbf{0805}, 107 (2008),
  [arXiv:0801.4921 [hep-th]].

\bibitem{Shen}
 J. Shen, B. Wang, C. Y. Lin, R. G. Cai, and R. K. Su,
{\em The phase transition and the quasi-normal modes of black holes},
  J. High Energy Phys. \textbf{0707}, 037 (2007),
  [arXiv:hep-th/0703102].

\bibitem{Rao}
X. P. Rao, B. Wang, and G. H. Yang,
  {\em Quasinormal modes and phase transition of black holes},
 Phys. Lett. B \textbf{649}, 472 (2007),
 [arXiv:0712.0645 [gr-qc]].

\bibitem{Jing}
 J. Jing and Q. Pan,
 {\em Quasinormal modes and second order thermodynamic phase transition for Reissner-Nordström black hole},
   Phys. Lett. B \textbf{660}, 13 (2008), [arXiv:0802.0043 [gr-qc]].

\bibitem{Berti}
 E. Berti and V. Cardoso,
 {\em Quasinormal modes and thermodynamic phase transitions},
   Phys. Rev. D \textbf{77}, 087501 (2008), [arXiv:0802.1889 [hep-th]].

\bibitem{He}
 X. He, S.-Chen, B. Wang, R.-G. Cai, and C.-Y. Lin,
 {\em Quasinormal modes in the background of charged Kaluza-Klein black hole with squashed horizons},
   Phys. Lett. B \textbf{665}, 392 (2008), [arXiv:0802.2449 [hep-th]].


\bibitem{He2}
 X. He, B. Wang, and S. Chen,
   {\em Quasinormal modes of charged squashed Kaluza-Klein black holes in the G?del Universe},
   Phys. Rev. D \textbf{79}, 084005 (2009),
 [arXiv:0811.2322 [gr-qc]].

\bibitem{LinLin}
 K. Lin, J. Li, and N. Yang,
   {\em Dynamical behavior and nonminimal derivative coupling scalar field of Reissner-Nordstroem black hole with a global monopole},
   Gen. Rel. Grav. \textbf{43} (2011).

\bibitem{Sup}
 Q.-Y. Pan and R.-K. Su,
   {\em Quasinormal Modes of Phantom Scalar Perturbation in Background of Reissner-Nordstrom Black Hole},
   Commun. Theor. Phys. \textbf{55}, 221 (2011).

\bibitem{Cardoso}
  V. Cardoso, A. S. Miranda, E. Berti, H. Witek, and V. T. Zanchin,
  {\em Geodesic stability, Lyapunov exponents and quasinormal modes},
   Phys. Rev. D \textbf{79}, 064016 (2009),
    	[arXiv:0812.1806 [hep-th]].

\bibitem{Stefanov}
  I. Z. Stefanov, S. S. Yazadjiev, and G. G. Gyulchev,
  {\em Connection between black-hole quasinormal modes and lensing in the strong deflection limit},
   Phys. Rev. Lett. \textbf{104}, 251103 (2010),
    	[arXiv:1003.1609 [gr-qc]].
	
\bibitem{Weic}
  S.-W. Wei and Y.-X. Liu,
  {\em Establishing a universal relation between gravitational waves and black hole lensing},
   Phys. Rev. D \textbf{89}, 047502 (2014),
    	[arXiv:1309.6375 [gr-qc]].

\bibitem{Kubiznak}
  D. Kubiznak and R. B. Mann,
  {\em $P$-$V$ criticality of charged AdS black holes},
   J. High Energy Phys. \textbf{1207}, 033 (2012),
    	[arXiv:1205.0559 [hep-th]].
	
\bibitem{Weia}
  S.-W. Wei and Y.-X. Liu,
  {\em Photon orbits and thermodynamic phase transition of $d$-dimensional charged AdS black holes},
   Phys. Rev. D \textbf{97}, 104027 (2018),
    	[arXiv:1711.01522 [gr-qc]].

\bibitem{Weib}
  S.-W. Wei, Y.-X. Liu, and Y.-Q. Wang,
  {\em Probing the relationship between the null geodesics and thermodynamic phase transition for rotating Kerr-AdS black holes},
  Phys. Rev. D \textbf{99}, 044013 (2019), [arXiv:1807.03455 [gr-qc]].	

\bibitem{Weid}
  Y.-M. Xu, H.-M. Wang, Y.-X. Liu, and S.-W. Wei,
  {\em Photon sphere and reentrant phase transition of charged Born-Infeld-AdS black holes},
  [arXiv:1906.03334 [gr-qc]].

\bibitem{Bhamidipati}
  C. Bhamidipati and S. Mohapatra,
  {\em Circular geodesics and phase transitions of charged black holes in AdS},
    	Phys. Lett. B \textbf{791}, 367 (2019),
    [arXiv:1805.05088 [hep-th]].

\bibitem{Goebel}
 C. J. Goebel,
   {\em Comments on the ``vibrations" of a black hole},
  Astrophys. J. \textbf{172}, L95 (1972).

\bibitem{Mashhoon}
 B. Mashhoon,
{\em Stability of charged rotating black holes in the eikonal approximation},
   Phys. Rev. D \textbf{31}, 290 (1985).

\bibitem{Iyer}
  S. Iyer,
  {\em Black-hole normal modes: A WKB approach. II. Schwarzschild black holes},
   Phys. Rev. D \textbf{35}, 3632 (1987).

\bibitem{Bertib}
  E. Berti and K. D. Kokkotas,
  {\em Quasinormal modes of Kerr-Newman black holes: coupling of electromagnetic and gravitational perturbations},
   Phys. Rev. D \textbf{71}, 124008 (2005),
    	[arXiv:gr-qc/0502065].

\bibitem{Decanini}
  Y. Decanini, A. Folacci, and B. Raffaelli,
  {\em Unstable circular null geodesics of static spherically symmetric black holes, Regge poles and quasinormal frequencies},
   Phys. Rev. D \textbf{81}, 104039 (2010),
    	[arXiv:1002.0121 [gr-qc]].

\bibitem{Hods}
  S. Hod,
  {\em The Reissner-Nordström black hole with the fastest relaxation rate},
   Eur. Phys. J. C \textbf{78}, 935 (2018),
    	[arXiv:1812.01014 [gr-qc]].

\bibitem{Cardosob}
  V. Cardoso, J. L. Costa, K. Destounis, P. Hintz, and A. Jansen,
  {\em Quasinormal modes and strong cosmic censorship},
   Phys. Rev. Lett. \textbf{120}, 031103 (2018),
    	[arXiv:1711.10502 [gr-qc]].

\bibitem{Hod}
  S. Hod,
  {\em Strong cosmic censorship in charged black-hole spacetimes: As strong as ever},
  Nucl.Phys. B \textbf{941}, 636 (2019),
    	[arXiv:1801.07261 [gr-qc]].

\bibitem{Dias}
  O. J. C. Dias, H. S. Reall, and J. E. Santos,
  {\em Strong cosmic censorship: taking the rough with the smooth},
  J. High Energy Phys. \textbf{1810}, 001 (2018),
    	[arXiv:1808.02895 [gr-qc]].

\bibitem{Cardosoc}
  V. Cardoso, J. L. Costa, K. Destounis, P. Hintz, and A. Jansen,
  {\em Strong cosmic censorship in charged black-hole spacetimes: still subtle},
  Phys. Rev. D \textbf{98}, 104007 (2018),
    	[arXiv:1808.03631 [gr-qc]].

\bibitem{Mo}
  Y. Mo, Y. Tian, B. Wang, H. Zhang, and Z. Zhong,
  {\em Strong cosmic censorship for the massless charged scalar field in the Reissner-Nordstrom-de Sitter spacetime},
  Phys. Rev. D \textbf{98}, 124025 (2018),
    	[arXiv:1808.03635 [gr-qc]].

\bibitem{Diasb}
  O. J. C. Dias, H. S. Reall, and J. E. Santos,
  {\em Strong cosmic censorship for charged de Sitter black holes with a charged scalar field},
  Class. Quant. Grav. \textbf{36}, 045005 (2019),
    	[arXiv:1808.04832 [gr-qc]].

\bibitem{Luna}
  R. Luna, M. Zilhao, V. Cardoso, J. L. Costa, and J. Natario,
  {\em Strong cosmic censorship: The nonlinear story},
  Phys. Rev. D \textbf{99}, 064014 (2019),
    	[arXiv:1810.00886 [gr-qc]].

\bibitem{Gej}
  B. Ge, J. Jiang, B. Wang, H. Zhang, and Z. Zhong,
  {\em Strong cosmic censorship for the massless Dirac field in the Reissner-Nordstrom-de Sitter spacetime},
  J. High Energ. Phys. \textbf{1901}, 123 (2019),
    	[arXiv:1810.12128 [gr-qc]].

\bibitem{Destounis}
  K. Destounis,
  {\em Charged Fermions and Strong Cosmic Censorship},
  Phys. Lett. B \textbf{795}, 211 (2019),
    	[arXiv:1811.10629 [gr-qc]].

\bibitem{LiuTang}
  H. Liu, Z. Tang, K. Destounis, B. Wang, E. Papantonopoulos, and H. Zhang,
  {\em Strong Cosmic Censorship in higher-dimensional Reissner-Nordstr\"{o}m-de Sitter spacetime},
  J. High Energ. Phys. \textbf{1903}, 187 (2019),
    	[arXiv:1902.01865 [gr-qc]].

\end{thebibliography}
\end{document}